# The Information Content of Sarbanes-Oxley in Predicting Security Breaches

J. Christopher Westland UIC

University of Illinois - Chicago

## Abstract

How effective is compliance with the Sarbanes-Oxley Act (SOX) in identifying and eliminating firm threats from information systems breaches?  We investigated publicly reported security breaches of internal controls in corporate systems to determine whether SOX assessments are information bearing with respect to breaches which can lead to materially significant losses and misstatements.   The results of our tests varied significantly between breach types.  SOX Section 404 adverse decisions on effectiveness of controls occurred in 100% of credit card data breaches and around 33% of insider breaches.  SOX 404 audits provided a contrarian "effective" control decisions on 88% of situations where there was a control breach concerning a portable device.  This suggests that employees are subverting particularly strict internal controls by using portable devices that can be carried outside the physical boundaries of the firm.  We found that management and SOX 404 auditors do not general agree on the underlying internal control situation at any time; instead the SOX 404 team was likely to discover material weaknesses and "educate" management and internal audit teams about the importance of these control weaknesses.  SOX attestations were poor at identifying control weaknesses from unintended disclosures, physical losses, hacking and malware, stationary devices, and situations where the cause of the breach was unknown.  Hazard and occupancy structural models were constructed to extrapolate to a larger population; results showed that both SOX 302 and 404 section audits provided information germane to the frequency of breaches, with SOX 404 being three times as informative as section 302 reports. The hazard model found an expected 2.88% reduction in breaches when SOX 302 controls are effective; management "material weakness' attestations provided no information in this structural model, whereas there would be around a 1% increase in breach occurrence when there are significant deficiencies. SOX 404 attestations were the most informative, and a negative SOX 404 attestation is projected to increase the frequency of breaches by around 8.5%. We concluded that the strength of internal controls attested in SOX reports is likely to be a significant factor in the occurrence of a breach at a firm in a specific period. *(349 words)*

1. The Sarbanes-Oxley Act and Threat to the Firm's Systems

Following the Enron and WorldCom collapses that marked the end of the dot-com bubble, U.S. legislators sought to better protect and inform investors through passage of the Sarbanes-Oxley Act of 2002 ("SOX"). Section 404 of SOX ("SOX 404") requires companies to review under the supervision of external auditors their *internal controls over financial reporting* ("ICFRs") and declare whether their ICFRs are "effective" or "ineffective." Section 302 of SOX ("SOX 302") requires companies to self-report on effectiveness of internal controls. One significant reason for the focus on internal control is the potential for firms to suffer systems intrusion from external actors, commonly referred to as security breaches or hacks, which can lead to materially significant losses and misstatements.

Recent information systems breaches have grown costlier and more frequent; for example, Home Depot's 2015 breach has cost it $232 million so far, an amount that they expect to reach billions. A 2015 breach of Ashley Madison stole 40 million accounts including photos, details of sexual proclivities and personal addresses, Target's 2013 breach affected the accounts of 70 million customers and so far, has cost the firm $162 million in added expense. In 2014 A "Guardians of Peace" breach of Sony Pictures stole over 100 terabytes of confidential data. 2014 also saw the theft of 360 million MySpace accounts, a LinkedIn hack that took more than 100 million accounts, a 500-million-account hack of Yahoo, 340 million AdultFriendFinder accounts (their second hack in a year) and numerous other breaches. Both frequency and scale of breaches have grown dramatically in the past five years.

Auditors have argued forcefully that they have no explicit responsibility for detecting fraud and external threats during audits. Nonetheless audits were forced to embrace some responsibility for detecting fraud and threats after the Enron and WorldCom collapses. (AICPA 2002) provides specific guidelines with respect to the auditor's responsibility for identifying external threats and fraud that may result in material misstatements. Security breaches and other external threats to the firm are subsumed under the category of "fraud" which AU 316.05 clarified.

> *" a broad legal concept and auditors do not make legal determinations of whether fraud has occurred. Rather, the auditor's interest specifically relates to acts that result in a material misstatement of the financial statements. The primary factor that distinguishes fraud from error is whether the underlying action that results in the misstatement of the financial statements is intentional or unintentional. Fraud is an intentional act that results in a material misstatement in financial statements that are the subject of an audit."*

The Sarbanes-Oxley Act was a formal attempt to impose additional joint responsibility on auditors and management for the detection of fraud and external threats, and for this research, SOX sections 302 and 404 which require internal control assessments.

SOX compliance has been contentious. Questions have been raised concerning the effectiveness of expensive SOX reviews, the external validity of SOX assessments for security breaches, errors, fraud and other external financial threats to the firm, and usefulness of SOX reporting to investors and other stakeholders. Unfortunately, these questions have been difficult to study, as external breaches, frauds and other crimes are significantly underreported in business, leading to a reporting bias in datasets that creates difficulties in controlled studies. Corporate executives, politicians and lobbyists have argued that the Sarbanes-Oxley Act (SOX) of 2002 is a cumbersome and costly regulation that is not effective (Drawbaugh 2012). Compliance with just section 404 has been estimated to average $1.7 million per firm annually (FEI 2007) and arguments have even been made before the US Supreme Court that SOX is unconstitutional (FEP 2010). Publications on

market and economy wide effects of SOX implementation have opined with case studies and polemics as well as empirical studies (Bratton 2003, Romano 2004, Coates 2007, Engel, Hayes et al. 2007, Kang, Liu et al. 2010) but the evidence on whether SOX implementation has improved security and integrity of internal controls is still equivocal.

This research addresses the question "How effective is corporate investment in SOX compliance in identifying and eliminating the threat from internal control security breaches?" External validity and utility of SOX assessments are difficult to measure because of the inclination for firms, conscious of their reputation, to significantly underreport internal control breaches after they happen. Effective breaches are seldom detected by the victim, and clearly cybercriminals themselves do not systematically report their activities. Nonetheless we believe that SOX metrics provide information on the potential for information systems security breaches since that is a fundamental objective of effective internal control over computer systems, and we attempt to provide empirical validation of these beliefs in this research.

Our current research investigates the information content of SOX reports with respect to security breaches to test five related hypotheses.

$H_1$: **SOX 404 attestations add new information to management's self-reported SOX 302 attestations**

*Importance*: SOX 404 attestations are costly and controversial, with some critics questioning whether they add information not already available in SOX 302 attestations.

$H_2$: **SOX 404 attests to the same set of system controls as SOX 302 attestations.**

*Importance*: Hypothesis $H_2$ complements hypothesis $H_1$ and tests whether SOX 302 and SOX 404 attestations relate to the same sets of controls and security issues.

$H_3$: **SOX 404 and 302 attestations contain information about the future occurrence of specific types of security breaches.**

*Importance*: This is our main hypothesis and tests will determine how much information SOX provides with respect to the various breach types.

$H_4$: **Structural models improve the accuracy of SOX 302 and SOX 404 attestations with respect to security breaches; they add information by positing an underlying breach generation model that fits empirically.**

*Importance*: Lucas' critique (Lucas 1976) suggested that structural models are needed to fully understand conclusions from regression analysis. We construct two models: (1) a hazard structural model and (2) an occupancy structural model to provide specific policy recommendations for internal control.

$H_5$: **Underlying control weaknesses cited in SOX attestations cause higher rates of security breaches, and not vice versa (counterfactuals)**

*Importance*: we provide a counterfactuals test to determine the direction of causality, and assure that SOX attestations are not simply citing control weaknesses after the fact because security breaches have occurred.

Our investigation proceeds as follows. Section 2 summarizes prior research that has investigated SOX reporting while characterizing SOX reporting following its full implementation in 2005. Section 3 reviews the data obtained for this research. Section 4 tests $H_1$ and $H_2$; section 5 tests $H_3$; section 6 tests $H_4$; and section 7 tests $H_5$. Section 8 summarizes these results with policy recommendations and recommendations for research which may extend the findings in this paper.

2. Prior Research on Information in SOX attestations

Most SOX studies to date have focused on internal consistency, compliance and accrual accounting assessments, rather than management of external risks such as fraud and security breaches. There should be a high correlation between accountants' assessment of control and their associated real-world consequences, but a variety of obstacles make external validity very difficult to study. Prior research has shown SOX 302 and SOX 404 metrics to be internally consistent within the bookkeeping and accounting domains. SOX 404 exceptions have been found to be correlated with financial restatements after discovery of real-world operational discrepancies and there is weak but positive support for SOX 404 assessments being consistent with good internal control (Rice and Weber 2012, Rice, Weber et al. 2014).

Compliance cost and information content of SOX filings with respect to other financial statistics reported by companies have been studied almost since the inception of SOX.  (Rice and Weber 2012, Rice, Weber et al. 2014) studied SOX 404 reporting for a sample of restating firms whose original misstatements are linked to underlying control weaknesses. They found that restatements, i.e. altering a prior year's financial statements due to auditor stakeholder or legal pressure, provided a great deal of information about control weaknesses that could lead to security breaches and management's awareness of these weaknesses.  In general, restating firms did acknowledge their existing control weaknesses during their misstatement periods and the probability of reporting existing weaknesses is negatively associated with external capital needs firm size non-audit fees and the presence of a large audit firm.  Restatements are positively correlated with financial distress, auditor effort, previously reported control weaknesses, other restatements and recent auditor or management changes.  (Rice and Weber 2012, Rice, Weber et al. 2014) found no evidence that penalties are more likely for firm managers or auditors that fail to report existing control weaknesses; rather they concluded that class action lawsuits management turnover and auditor turnover are all more likely in the wake of a restatement when control weaknesses had previously been reported.  They concluded that existing public and private enforcement mechanisms surrounding SOX 404 are unlikely to provide strong incentives for compliance and offer a potential explanation for why most restatements are issued by firms that previously claimed to have effective internal controls.

 (Bedard, Hoitash et al. 2009, Bedard and Graham 2011) examined detection and severity classification of internal control deficiencies, finding that external auditors during their Section 404 audit detect about three-fourths of unremediated internal control deficiencies.  (Ge, Koester et al. 2016) looked at a sample of 261 companies that disclosed at least one material weakness in internal control in their SOX filings, finding that poor internal control is usually related to an insufficient commitment of resources for accounting controls, with the most common account-specific material weaknesses occurring in accounts receivable and inventory.  SOX 302 disclosures, in contrast, tended to describe internal control problems in complex accounts such as the derivative and income tax accounts. They found that disclosing a material weakness is positively associated with business complexity (e.g. multiple segments and foreign currency), negatively associated with firm size (e.g. market capitalization) and negatively associated with firm profitability (e.g. return on assets). (Lin, Pizzini et al. 2011) investigated the role that a firm's internal audit function plays in the disclosure of material weaknesses reported under SOX 404 using data from 214 firms. They found that material weakness disclosures are negatively correlated with the education level of the internal auditors and positively correlated with the practice of grading audit engagements and external-internal auditor coordination. (Ashbaugh-Skaife, Collins et al. 2007) reported that SOX disclosed internal control deficiencies were associated with more complex operations, recent organizational changes, greater accounting risk, more auditor resignations and have fewer resources available for internal control. They also found that firms with SOX disclosed internal control deficiencies had more prior SEC enforcement actions and financial restatements, were more likely to use a single dominant audit firm, and had more concentrated institutional ownership. (Feng, Li et al. 2009) found that internal control deficiencies were correlated with less accurate guidance. In particular, the impact of ineffective internal controls on forecast accuracy was found to be three times larger when the weakness related to revenues or cost of goods sold because of the importance of these two accounts in forecasting earnings. (Ashbaugh-Skaife, Collins et al. 2008) found that firms reporting internal control deficiencies have lower quality accruals as measured by accrual noise and absolute abnormal accruals when compared to firms not reporting internal control problems. Additionally, firms whose auditors confirm remediation of previously reported internal control deficiencies exhibit an increase in accrual quality relative to firms that do not remediate their control problems. They found that material weaknesses are correlated with noise to accrual 'noise' (higher error term variance) and intentional misstatements that bias earnings upward.

### 3. Breach and SOX Data Items

SOX reporting began in 2004 but only in 2005 did all publicly traded companies report their SOX assessments. This research collected SOX annual reports as well as publicly reported security breaches in the 10-year period 2005-2015. We considered two overlapping groups of firms in our study: (1) publicly traded firms that reported SOX assessments and (2) publicly traded firms that had reported security breaches in the research period. Figure 2 depicts the overlapping sets of: (A) all US firms, (B) publicly traded US firms, and (C) public and private US firms that have reported security breaches. We ignored management-only SOX assessments which primarily apply to small private firms and focused our research on the more complete dataset of publicly traded companies *that includes external audit assessments of controls* (B in figure 2).

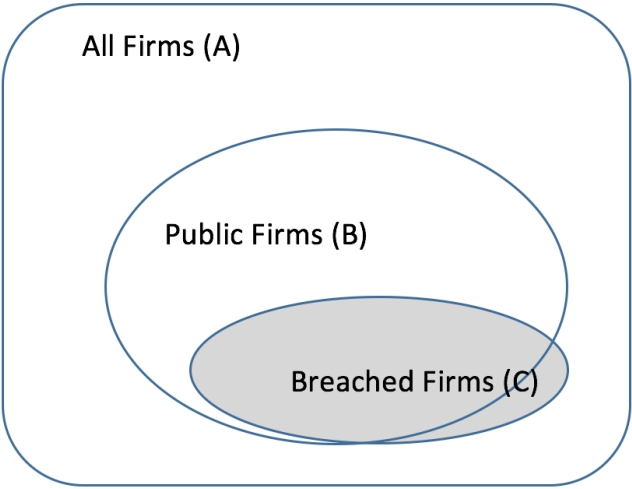

*Figure 1: Four subsets of firms for which data was obtained in this research*

We defined two additional subsets $B \cap C$ (public firms that experienced a breach) and $B \cap \neg C$ (public firms that did not experience a breach) where we proceeded with separate customized analyses to control selection bias. To build the data subsets $B \cap C$ and $B \cap \neg C$ it was necessary to join the information on SOX and Breach reporting on a $[firm - date]$ index. The data is not directly comparable between these two subsets since SOX reports are periodic filings by management and auditors that are conditioned on managerial contracting processes *not* internal control processes. In contrast the Breach dataset provides reports on the breakdown of internal control processes and are incomplete because companies and criminals do not like to report breaches; industry estimates suggest that between 3% and 30% of breaches are publicly reported (Menn 2012). SOX reports will be generated annually whether anything occurs or changes with respect to internal control; but breaches clearly require some sort of control failure. Table 1 summarizes the data items extracted from the SOX 302 and 404 reporting that were used in the analysis.

*Table 1: SOX 302 and 404 Metrics used in the Analysis*

| | **SOX 302 metrics:** | |
|---|---|---|
| IS_EFFECTIVE (302) | Self-reported management assertion concerning whether internal controls are effective in the firm. | 0 = no<br>1 = yes |
| MATERIAL_WEAKNESS (302) | Self-reported management assertion concerning whether there exist material weaknesses in internal control systems in the company. | 0 = controls are effective<br>1 = there are material weaknesses in the firm's controls |
| SIG_DEFICIENCY (302) | Self-reported management assertion concerning whether there exist significant deficiencies in internal control systems in the company | 0 = controls are effective<br>1 = there are deficiencies in the firm's controls |
| | **SOX 404 metrics:** | |
| IC_OP_TYPE | Manual system or Automated computer system | 0 = Manual<br>1 = Automated |
| AUDITOR_AGREES (404) | Auditors 'attestation' concerning management's SOX 404 assertions | 0 = disagree<br>1 = agrees |
| COMBINED_IC_OP (404) | Combined opinion on the effectiveness of the firm's internal controls | 0 = controls are generally effective<br>1 = there exists a significant control deficiency |
| IC_IS_EFFECTIVE (404) | Management asserts that internal controls are effective (Internal Control is Effective) | 0 = no<br>1 = yes |
| AUDIT_FEE | Cost of SOX 404 audit | (numeric; monetary) |

This research collected 170647 annual SOX 302 and SOX 404 SEC 10-K 10-KSB 10-Q 10-QSB 20-F and 40-F filings of U.S. firms publicly listed on the NYSE or NASDAQ (Firms began to report this data starting April 15 2005 but advanced filings appeared as early as 2002). Out of these we selected the annual reporting only since these had the audited SOX information (Section 404) as well as a comparable set of management reported SOX information (Section 302). We also acquired the Privacy Rights Clearinghouse / Verizon compilation of 388 major reported U.S. corporate security breaches since 2005 of which 213 applied to 184 NYSE and NASDAQ firms.

*Table 2: Breach Data used in the analysis*

| | Count |
|---|---|
| SOX Reports: | |
| • NASDAQ firms | 1835 |
| • NYSE firms | 1563 |
| Total Firms | **3398** |
| • SOX Reports on Automated Systems | 27496 |
| • SOX Reports on Manual Systems | 31441 |
| Total SOX reports in NASDAQ & NYSE (dataset B) | 58937 |
| | |
| Breach Data: | |
| • Payment Card Fraud (CARD) - Fraud involving debit and credit cards that is not accomplished via hacking. For example skimming devices at point-of-service terminals | 12 |
| • Unintended disclosure (DISC) - Sensitive information posted publicly on a website mishandled or sent to the wrong party via email fax or mail. | 70 |
| • Hacking or malware (HACK) - Electronic entry by an outside party malware and spyware. | 106 |
| • Insider (INSD) - Someone with legitimate access intentionally breaches information - such as an employee or contractor. | 51 |
| • Physical loss (PHYS) - Lost discarded or stolen non-electronic records such as paper documents | 20 |
| • Portable device (PORT) - Lost discarded or stolen laptop PDA smartphone portable memory device CD hard drive data tape and so forth. | 103 |
| • Stationary device (STAT) - Lost discarded or stolen stationary electronic device such as a computer or server not designed for mobility. | 11 |
| • Unknown (UNKN) | 15 |
| • Total number of breaches reported from 2005 to 2016 (dataset C) | <u>388</u> |
| Total number of breaches applicable to NYSE and NASDAQ listed firms (dataset $B \cap C$) | 213 |
| Unique firms in breach dataset listed on NYSE & NASDAQ | 184 |

Industry sources suggest that security breaches are substantially underreported by factors ranging from 3 to 30 (Menn 2012, Symantic 2016). Cybercrime is a low-risk high return vocation that (Lewis and Baker 2013) estimate annually costs the global economy between $375 to $575 billion in losses. Most breaches are not reported as they may signal to customers and investors a lack of internal control. This is especially true with financial institutions such as banks and insurance companies, who can avoid some of the public scrutiny of customer facing retail and Internet firms. In addition, the best hackers are often state-sponsored security professionals, using methods that are undetectable by current technologies and can subvert internal controls and compromise assets without ever being detected. Breaches, fraud, compromised computer systems as well as SOX compliance concerns have motivated billions of dollars of expenditure on computers and information systems upgrades and replacements.

SOX statistics show that the adverse percentage rate for auditor attestations decreased rapidly from a high in 2004 of 16.9% to 10.3% in 2005 and by 2009 had leveled off at around 2% indicating either (1) management and internal staff were increasingly adept at convincing external auditors that their systems were secure and protected against fraud and external threats; or (2) there was a significant improvement in internal control immediately after implementation of SOX, at least in those areas that were included in the SOX audits. There were additional "management-only" assessments primarily apply to small private firms; although trends for these firms roughly followed those of larger firms, the rate of issuance of adverse opinions typically was around 30% higher than for publicly traded companies (Cheffers 2012) and the percentage of adverse opinions has declined in tandem with those of publicly traded firms.

In addition, each of the "Big 4" audit firms (which audit almost all of the listed firms on US exchanges) displayed unique biases in rendering adverse attestations: Ernst & Young focused on accounts receivables, revenue recognition, taxes and fixed assets; PricewaterhouseCoopers focused on accounts receivables, revenue recognition, taxes and payables; KPMG focused on accounts receivables, revenue recognition, taxes and inventory; and Deloitte & Touche focused on revenue recognition, taxes, liabilities, inventory and executive compensation (Cheffers 2012). These biases are likely to reflect signature audit methods, internal forms and checklists, and audit histories that are unique to individual firms. These firms will consequently allocate larger portions of the audit budget to certain accounts at the expense of others. Additionally, auditors tend to allocate more time to auditing debit balance accounts, assuming double entry will assure the accuracy of the credit accounts; but the accounts selected for audit depend on firm policy, procedures and managing partners.

Information systems security vulnerabilities are addressed in auditors' assertions of whether specific "material weaknesses" have been found. These vary from year to year as old weaknesses are addressed and new weaknesses surface, but eight financial reporting areas that were cited as poorly controlled in an adverse SOX Opinion: (1) Accounts/Loans Receivable (2) Investments or Cash; (3) Securities (Debt Quasi-Debt Warrants and Equities); Revenue Recognition; (4) Tax Expense/Benefit/Deferral (FAS109); (5) Liabilities Payables Reserves and Accrual Estimates; (6) Entity (Foreign Related Party Affiliated Subsidiary); (7) PPE Intangibles Fixed Assets; and (8) Inventory Vendor Cost of Sales. Only (1), (2) and (8) tend to be control areas that will show up in reported security breaches. Additionally eight operational areas commonly appeared in SOX 404 attestations: (1) response to prior SOX 404 opinions; (2) response to material weakness; (3) personnel issues; (4) segregation of duties; (5) restatements of financials; (6) material year-end adjustments; (7) internal audit findings; and (8) IT Systems. Only (3), (4) and (8) tend to be operational areas that will show up in reported security breaches, although (Rice and Weber 2012, Rice, Weber et al. 2014) found that (5), the restatement of financials, was a significant indicator of internal control weaknesses that would lead to adverse SOX 302 and 404 assessments and potential security weaknesses.

The basic "unit" of observation in this analysis is the publication of a SOX report and breaches are interpreted as *ex post* evidence concerning information in these SOX reports. Whether *ex post* means correlation or causality will be explored in our test of $H_6$ counterfactuals. The breach information is both censored (it has only been collected between 2005 and 2015) and it is incomplete as only 3% to 30% of breaches are publicly reported on this dataset (Menn 2012). This is a significant problem in inference as it suggests that the observed (reported) breaches are a subset of the population. In randomized samples 3-30% would be sufficient for inference; and field studies commonly appeal to the concept of a "convenience sample." We judged it likely to find some systematic biases in both occurrence and reporting of breaches with consumer facing industries more likely to report (because it is difficult to hide a breach from customers) and back office financial industries less likely to report due to the impact on their reputation for security and efficiency. We address these potential biases in the structural models presented later in the paper but initially construct Breach-SOX panel data for regression incorporating available data.

In order to merge the two datasets we organized them into observations indexed on $\{firm\ identifier\} + \{date\}$. Neither the set of firms nor the dates correspond in the two datasets. SOX reporting dates occur annually for publicly listed firms and it is sufficient simply to index them by year. Breach reports though appear in a press release or news article whose date has daily resolution and can include both listed and unlisted firms. We have expanded the breach event period to on year, since our dataset includes only a report date, and the breach itself often spanned a period, usually not longer than one year. With these generalizations, it is possible to merge the breach and SOX data indexed on $\{firm\ identifier\} + \{year\}$.

The dataset reports breaches for 213 of the 3398 listed firms which report SOX assessments between 2005 and 2015. Restriction of the analysis to subsets $B \cap C$ in figure 2 (public firms that experienced a breach) resolves the issue with firms by dropping the breached firms for which we have no SOX 404 reports and a conceptual model for the panel regression $(breaches \sim SOX)_{B \cap C}$ where $breaches$ represent security breach classifications and other metrics from the Privacy Clearinghouse / Verizon dataset and $SOX$ represents SOX 302 and 404 reported assessments from the AuditAnalytics dataset. Datasets were dichotomized into manual and automated systems each which receives separate SOX reports.

### 4. Tests of SOX Information Content

The rapid initial decline in the SOX adverse percentage rate for auditor attestations from an initial 16.9% to a current level of around 2% suggests that SOX 404 audit procedures, which are estimated to cost around $2 million annually, are not adding significant information about systems security over that provided by management in internal auditors at the firm in their self-reported SOX 302 attestations. To investigate, we tested $H_1$: SOX 404 attestations add new information to management's self-reported SOX 302 attestations. This will test the veracity of critiques that SOX 404 attestations are costly without providing new information not already available in SOX 302 attestations.

We concluded that any confounding SOX attestations effects due to firm size were insignificant with respect to the research questions investigated here. This was not unexpected as neither (Ge, Koester et al. 2016) nor (Ashbaugh-Skaife, Collins et al. 2007) discovered size effects in SOX 404 disclosures, though they did find that firm complexity did contribute to the number of adverse attestations under section 404. To assess the possibility of size effects confounding our main tests about information in SOX attestations about security breaches, four measures of firm size – sales, net income, market capitalization, and number of employees – were regressed against breaches.

Table 3 summarizes the results: the four $R^2$s are around 1-5% and the only significant relationship we found was that firms with more employees have more "insider" breaches (INSD) which seems a logical consequence of a larger more labor-intensive operations. Other financial metrics are only weakly correlated with all types of security breaches. An insignificant amount of variance in breach occurrences depends on structural parameters and fit statistics $t-statistics$, $R^2$'s and $F-statistics$ indicate that firm size is an insignificant influence on the rate of occurrence of security breaches.

*Table 3: OLS Regression of Breach Type Occurrence against Firm Size and Operations Metrics*

|  | Capitalization | | | Sales | | | Income | | | Number of employees | | |
|---|---|---|---|---|---|---|---|---|---|---|---|---|
| y → <br> x ↓ | Estimate | t | Sig | Estimate | t | Sig | Estimate | t | Sig | Estimate | t | Sig |
|  | 36555 | 4.655 | *** | 29591 | 6.244 | *** | 2664.76 | 3.279 | *** | 74.164 | 6.029 | *** |
| CARD | 5973 | 0.591 |  | 7754 | 1.167 |  | 396.35 | 0.348 |  | 20.796 | 1.214 |  |
| DISC | -10479 | -1.228 |  | 11702 | 2.194 | * | -72.33 | -0.079 |  | 26.028 | 1.891 | . |
| HACK | -14291 | -1.697 | . | -11826 | -2.205 | * | -1432.22 | -1.557 |  | -11.49 | -0.83 |  |
| INSD | -6047 | -0.698 |  | 13329 | 2.377 | * | -661.13 | -0.688 |  | 64.556 | 4.474 | *** |
| PHYS | -6043 | -0.576 |  | 4949 | 0.714 |  | -913.76 | -0.769 |  | -6.953 | -0.39 |  |
| PORT | 13050 | 1.582 |  | 12670 | 2.475 | * | 162.26 | 0.185 | * | 32.534 | 2.459 | * |
| STAT | -22686 | -2.298 | * | 7116 | 1.045 |  | -2604.27 | -2.23 |  | -8.22 | -0.473 |  |
| UNKN | NA | NA |  | NA | NA |  | NA | NA |  | NA | NA |  |
| $R^2$ | 0.05429 |  |  | 0.04479 |  |  | 0.01104 |  |  | 0.04741 |  |  |
| F | 10.43 |  |  | 10.74 |  |  | 2.556 |  |  | 11.28 |  |  |

Signif. codes: 0 '***' 0.001 '**' 0.01 '*' 0.05 '.' 0.1 ' ' 1

We tested panel data of annual firm reports (subset B in figure 2) from 2005 through 2015 separately for audits of manual and automated system with conceptual model[1] $(SOX404 \sim SOX302)_B$ which was implemented in the following regression model:

---

[1] For conceptual discussions, I have throughout this paper used the R-language format of "response variable ~ predictor variables" where the tilde means "is modeled as a function of." The parentheses and subscript further define the population in which the model is tested.

$$\text{SOX404}_{\text{IC}} = \beta_0 + \beta_1 \text{SOX302}_{\text{IC}} + \beta_2 \text{SOX302}_{\text{MW}} + \beta_3 \text{SOX302}_{\text{SD}} +$$
$$+\beta_4 \text{SOX302}_{\text{Audit}} + \beta_5 \text{SOX302}_{\text{NA}} + \beta_6 \text{SOX302}_{\text{Tax}} + \beta_7 \text{SOX302}_{\text{Related}} + \beta_8 \text{SOX302}_{\text{Other}} \quad (1)$$

Where:

| Variable | Description |
|---|---|
| $SOX404_{IC}$ | Internal Controls are Effective Indicator Variable |
| $SOX302_{IC}$ | Internal Controls are Effective Indicator Variable |
| $SOX302_{MW}$ | Material Weaknesses Exist Indicator Variable |
| $SOX302_{SD}$ | Significant Deficiencies Exist Indicator Variable |
| $SOX302_{Audit}$ | Audit Firm Audit Fees |
| $SOX302_{NA}$ | Non-Audit Firm Fees |
| $SOX302_{Tax}$ | Audit Firm Tax Fees |
| $SOX302_{Related}$ | Audit Firm Related Fees |
| $SOX302_{Other}$ | Audit Firm Other Fees |

Wooldridge's test for unobserved individual effects yielded z = 2.535 (p-value = 0.01124) for the automated systems audits; and for manual systems z = 3.7439 (p-value = 0.0001812) both supporting a hypothesis that unobserved effects exist, indicating we need to use fixed effects regression. Wooldridge's test for serial correlation in fixed effects panels yielded $\psi^2$ = 0.24677 (p-value = 0.6194) and for manual systems $\psi^2$ = 4.1434 (p-value = 0.0418) both suggesting that serial correlation exists in the panels. Unobserved effects and serial correlation would be expected if we consider a structural analysis of auditing. Audit scope is limited by regulation, so it would be expected that there are unobserved effects that are not covered by this regulation. Firms that have poor security in one year, are likely to have poor security in other years, and thus the indicators and fees will both exhibit serial correlation. Tables 4 and 5 summarize the panel regression results.

*Table 4: Fixed Effects Panel Regression estimates for SOX 404 "Internal Controls are Effective in the Firm's **Automated** Systems" Decisions (dependent) on SOX 302 and Management Decisions (independent)*

|  | Estimate | Std. Error | t-value | Pr(>|t|) |  |
|---:|---:|---:|---:|---:|---|
| IS_EFFECTIVE | 0.155 | 0.004 | 40.606 | 0.000 | *** |
| MATERIAL_WEAKNESS | -0.810 | 0.004 | -198.815 | 0.000 | *** |
| SIG_DEFICIENCY | -0.009 | 0.001 | -7.047 | 0.000 | *** |
| AUDIT_FEES | -1.41E-10 | 9.60E-11 | -1.4715 | 0.1412 | |
| NON_AUDIT_FEES | 2.77E-08 | 2.95E-08 | 0.9396 | 0.3474 | |
| TAX_FEES | -2.70E-08 | 2.95E-08 | -0.9164 | 0.3595 | |
| AUDIT_RELATED_FEES | -2.76E-08 | 2.95E-08 | -0.9367 | 0.3489 | |
| $R^2$ | 0.90782 | | | | |
| F-statistic | 29507.7 | | | | |
| Residual Plot | | | | | |

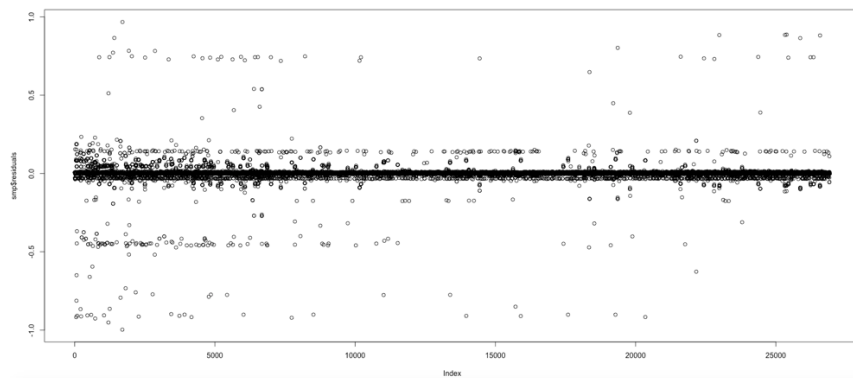

In the reports of both manual and automated systems, the estimators indicate that firms on average attest that internal controls are effective six times $\left(6 \cong {1}/{0.155} \cong {1}/{0.175}\right)$ as often in SOX 302 reports as in SOX 404 reports. The signs for all the variables are in the expected direction and there was no evidence that any of the fees charged to the firm influenced the auditor's opinion – we validated the independence of SOX auditors. In both manual and automated systems, the firm's self-reported "material weaknesses" were associated with around 80% of SOX 404 negative control assessments, while self-reported "significant deficiencies" appeared in only around 1-2% of SOX 404 negative control assessments. This suggests that management's self-reported "material weaknesses" were a significant indicator of control problems identified by auditors, while significant deficiencies were not.

*Table 5: Fixed Effects Panel Regression estimates for SOX 404 "Internal Controls are Effective in the Firm's **Manual** Systems" Decisions (dependent) on SOX 302 and Management Decisions (independent)*

|  | Estimate | Std. Error | t-value | Pr(>\|t\|) |  |
|---|---|---|---|---|---|
| IS_EFFECTIVE | 0.175 | 0.004 | 40.080 | 0.000 | *** |
| MATERIAL_WEAKNESS | -0.796 | 0.005 | -173.302 | 0.000 | *** |
| SIG_DEFICIENCY | -0.016 | 0.002 | -10.241 | 0.000 | *** |
| AUDIT_FEES | 1.33E-10 | 1.26E-10 | 1.0608 | 0.2888 |  |
| NON_AUDIT_FEES | 5.37E-08 | 3.86E-08 | 1.3911 | 0.1642 |  |
| TAX_FEES | -5.28E-08 | 3.86E-08 | -1.3681 | 0.1713 |  |
| AUDIT_RELATED_FEES | -5.39E-08 | 3.86E-08 | -1.3973 | 0.1623 |  |
| $R^2$ | 0.86641 |  |  |  |  |
| F-statistic | 22251.6 |  |  |  |  |
| Residual Plot | | | | | |

The correlation between an auditors' SOX404 attestation in a year is strongly (99.99% confidence) dependent on $SOX302_{IC}$, $SOX302_{SD}$ and $SOX302_{Audit}$ and slightly less so for $SOX302_{MW}$. The amounts of variance explained in the automated and manual systems data subsets are $R^2 = 91\%$ and $R^2 = 87\%$ respectively, which lends credence to the parameter estimates. Our findings strongly support the hypothesis $H_2$ that SOX 404 testing adds new information to self-reported SOX 302 attestations, and suggest that SOX 404 audits, though influenced by management's self-reported "material weaknesses," are additionally providing investors with information not otherwise available to stakeholders.

Serial correlation in the panel data was detected via Wooldridge's test, and we have taken advantage of this finding by removing this effect in testing $H_3$ by summing (aggregating) all the data by firm. Any contemporaneous agreement by auditors and management should be washed out in the aggregation over time. Any remaining questions concerning auditors' independence in the conduct of SOX 404 tests could possibly have arisen because underlying control weaknesses are obvious, and do not necessarily require independent attestation to identify. Appearances of control weaknesses are expected to be transient and time dependent, i.e., management is strongly inclined to rectify them once they are discovered, and thus the same weaknesses should not be present in subsequent time periods. If we test the previous model with data that has been aggregated over the time periods from 2005 to 2015, we will smooth any transient effects and reveal sustained biases and lack of independence in SOX 404 attestations. To this end we tested $H_2$: SOX 404 attests to controls over the same systems as management and internal auditors SOX 302 attestations, to determine if the convergence of opinions of auditors and management are due the same set of underlying control weaknesses that exist at a point in time.

Because we chose a fixed effects model for the panel regressions, we have chosen to retain the intercept in the regressions of the aggregated data to absorb any unobserved effects, making the results between the two regressions comparable. The results presented in tables 6 and 7 and estimators have changed substantially because of regression to the mean.

*Table 6: Regression estimates for SOX 404 "Internal Controls are Effective in the Firm's **Automated** Systems" Decisions (dependent) on SOX 302 and Management Decisions (independent) aggregated over years 2005 to 2015*

|  | Estimate | Std. Error | t-value | Pr(>|t|) |  |
|---:|---:|---:|---:|---:|---|
| (Intercept) | 0.381700 | 0.038310 | 9.964000 | 0.000000 | *** |
| IS_EFFECTIVE | 1.970000 | 0.003912 | 503.603000 | 0.000000 | *** |
| MATERIAL_WEAKNESS | 0.893900 | 0.016250 | 55.011000 | 0.000000 | *** |
| SIG_DEFICIENCY | -0.037210 | 0.010710 | -3.474000 | 0.000521 | *** |
| AUDIT_FEES | 0.000001 | 0.000000 | 3.252000 | 0.001157 | ** |
| NON_AUDIT_FEES | 0.000404 | 0.000141 | 2.869000 | 0.004143 | ** |
| TAX_FEES | -0.000401 | 0.000141 | -2.848000 | 0.004427 | ** |
| AUDIT_RELATED_FEES | -0.000407 | 0.000141 | -2.892000 | 0.003853 | ** |
| OTHER_FEES | -0.000403 | 0.000141 | -2.866000 | 0.004192 | ** |
| $R^2$ | 0.9906 |  |  |  |  |
| F-statistic | 38700 |  |  |  |  |

*Table 7: Regression estimates for SOX 404 "Internal Controls are Effective in the Firm's **Manual** Systems" Decisions (dependent) on SOX 302 and Management Decisions (independent) aggregated over years 2005 to 2015*

|  | Estimate | Std. Error | t-value | Pr(>\|t\|) |  |
|---:|---:|---:|---:|---:|---|
| (Intercept) | 0.611400 | 0.060790 | 10.059000 | 0.000000 | *** |
| IS_EFFECTIVE | 2.956000 | 0.006189 | 477.640000 | 0.000000 | *** |
| MATERIAL_WEAKNESS | 1.763000 | 0.022310 | 79.031000 | 0.000000 | *** |
| SIG_DEFICIENCY | -0.052900 | 0.016950 | -3.120000 | 0.001822 | ** |
| AUDIT_FEES | 0.000002 | 0.000001 | 3.641000 | 0.000276 | *** |
| NON_AUDIT_FEES | 0.000590 | 0.000235 | 2.513000 | 0.012017 | * |
| TAX_FEES | -0.000585 | 0.000235 | -2.495000 | 0.012660 | * |
| AUDIT_RELATED_FEES | -0.000595 | 0.000235 | -2.536000 | 0.011243 | * |
| OTHER_FEES | -0.000590 | 0.000235 | -2.515000 | 0.011947 | * |
| $R^2$ | 0.988 | | | | |
| F-statistic | 34480.00 | | | | |

In the reports of both manual and automated systems, the estimators indicate that firms on average attest that internal controls are effective one-third to one-half $\left(0.3 \text{ to } 0.5 \cong {}^1/_{2.96} \text{ to } {}^1/_{1.97}\right)$ as often in SOX 302 reports as in SOX 404 reports. Overall, firms seem to be substantially more vigilant and suspicious of internal control weaknesses than SOX 404 auditors. It is likely that management and internal auditors have more information about control weaknesses, since they audit the firm year-round, as opposed to the annual engagements of SOX404 teams. The amounts of variance explained in the automated and manual systems data subsets are both $R^2 \cong 99\%$ and this improvement probably reflects the regression to the mean because of aggregation. The impact of "significant deficiency" assertions in SOX 302 appear consistent with the panel regressions.

Magnitude changes on the SOX 302 "control is effective" and "material weakness" estimators as well as the sign reversals "material weakness" estimates lead us to reject $H_3$. The sign reversal from panel regression results of the "material weakness" estimates suggests that the absence of "material weaknesses" self-reported by management in the SOX 302 reports was associated with significantly higher incidence of negative SOX 404 control assessments in general. This is consistent with prior research by (Benoit 2006, Bedard, Hoitash et al. 2009) who found that management is overconfident in assessing internal control effectiveness reflected in self-reporting under SOX302 where 90% of companies with ineffective Section 404 controls (IC_IS_EFFECTIVE = 0) still self-reported effective 302 controls (IS_EFFECTIVE=1) in the same period end that an adverse Section 404 was reported. We confirmed Benoit's findings: in our data Welch's Two Sample t-statistic was $-0.24999$ $(p-val: 0.8026)$. This may reflect that management is generally unaware of material weaknesses that the SOX 404 external auditors deem important, and that auditors "educate" management about the material weaknesses during the conduct of their SOX 404 attestations. We concluded that rather than both management and auditors discovering the same underlying internal control situation at any time; instead the SOX 404 auditors were likely to discover material weaknesses and "educate" management and internal audit teams about the importance of these control weaknesses.

### 5. Predicting Breaches

Our tests to this point have shown that the occurrence of breaches is only weakly affected by size, and that the strength of internal controls is likely to be a significant factor in the occurrence of a breach at a firm in a period. We also concluded that SOX 404 auditors maintain their independence consistent with Generally Accepted Auditing Standard (GAAS). With the investigation of potentially confounding factors of size and auditor independence effectively completed, we can move on to testing our main hypothesis $H_3$: SOX 404 and 302 attestations contain information about the future occurrence of specific types of security breaches.

We tested panel data of annual firm reports (intersection $B \cap C$ in figure 2) from 2005 through 2015 separately for audits of manual and automated system with conceptual model $(breaches \sim SOX)_{B \cap C}$ which was implemented in the following regression model:

$$breach = \beta_0 + \beta_1 SOX302_{IC} + \beta_2 SOX302_{MW} + \beta_3 SOX302_{SD} + \\ +\beta_4 SOX404_{IC} + +\beta_5 SOX302_{Audit} + \beta_6 SOX302_{NA} + \beta_7 SOX302_{Tax} + \\ + \beta_8 SOX302_{Related} + \beta_9 SOX302_{Other} \quad (2)$$

|  | Estimators | | | | | | $R^2$ | F | Woolridge Individual Effects Test | | Woolridge Serial Correlation Test | |
|---|---|---|---|---|---|---|---|---|---|---|---|---|
|  | SOX 404 Effective Control | SOX 302 Effective Control | SOX 302 Mtl Weakness | SOX 302 Sig Deficiency | Audit Fees | Other Fees |  |  | z | p-value | $\psi^2$ | p-value |
| Payment Card Fraud (CARD) | -0.9953 | 0.9913 | 0.0071 | -0.0238 | 0.0000 | 0.0020 | 0.1681 | 1.8182 | -0.9029 | 0.3666 | 4.3 | 0.0376 |
| (t-value) | -2.8160 | 2.5910 | 0.0515 | -0.4932 | -0.5547 | 0.1638 |  |  |  |  |  |  |
| (p-value) | 0.0061 | 0.0113 | 0.9591 | 0.6232 | 0.5807 | 0.8703 |  |  |  |  |  |  |
| Unintended disclosure (DISC) | 0.1417 | 0.0452 | 0.4832 | -0.1052 | 0.0000 | 0.0046 | 0.2411 | 2.8594 | 1.6187 | 0.1055 | 11.8 | 0.0006 |
| (t-value) | 0.1788 | 0.0527 | 1.5623 | -0.9711 | -0.3400 | 0.1676 |  |  |  |  |  |  |
| (p-value) | 0.8585 | 0.9581 | 0.1221 | 0.3344 | 0.7347 | 0.8673 |  |  |  |  |  |  |
| Hacking or malware (HACK) . | 0.4715 | -0.4426 | -0.3179 | 0.0541 | 0.0000 | -0.0042 | 0.0400 | 0.3752 | 1.4735 | 0.1406 | 4.8 | 4.7971 |
| (t-value) | 0.5630 | -0.4883 | -0.9724 | 0.4728 | 0.2287 | -0.1441 |  |  |  |  |  |  |
| (p-value) | 0.5750 | 0.6267 | 0.3337 | 0.6376 | 0.8197 | 0.8858 |  |  |  |  |  |  |
| Insider (INSD). | -0.3209 | 0.5343 | -0.1124 | 0.1895 | 0.0000 | -0.0214 | 0.2234 | 2.5882 | 1.9584 | 0.0502 | 11.0 | 0.0009 |
| (t-value) | -0.4339 | 0.6674 | -0.3893 | 1.8743 | -1.4343 | -0.8355 |  |  |  |  |  |  |
| (p-value) | 0.6656 | 0.5064 | 0.6981 | 0.0645 | 0.1554 | 0.4059 |  |  |  |  |  |  |
| Physical loss (PHYS) | 0.1981 | 0.0660 | 0.3679 | 0.0528 | 0.0000 | 0.0461 | 0.3592 | 5.0440 | 0.6085 | 0.5428 | 130.3 | 0.0000 |
| (t-value) | 0.3718 | 0.1144 | 1.7681 | 0.7249 | 0.0191 | 2.4913 |  |  |  |  |  |  |
| (p-value) | 0.7110 | 0.9092 | 0.0808 | 0.4706 | 0.9848 | 0.0148 |  |  |  |  |  |  |
| Portable device (PORT) | 0.8772 | -0.6878 | 0.4239 | -0.1119 | 0.0000 | -0.0235 | 0.1739 | 1.8943 | 0.5024 | 0.6154 | 3.0 | 0.0851 |
| (t-value) | 0.8690 | -0.6294 | 1.0755 | -0.8104 | 1.6993 | -0.6716 |  |  |  |  |  |  |
| (p-value) | 0.3874 | 0.5309 | 0.2853 | 0.4201 | 0.0931 | 0.5037 |  |  |  |  |  |  |
| Stationary device (STAT) | 0.0480 | -0.0590 | -0.0150 | 0.0011 | 0.0000 | 0.0002 | 0.0015 | 0.0135 | 1.4913 | 0.1359 | 828280.0 | 0.0000 |
| (t-value) | 0.1807 | -0.2053 | -0.1443 | 0.0304 | 0.3264 | 0.0223 |  |  |  |  |  |  |
| (p-value) | 0.8571 | 0.8379 | 0.8856 | 0.9759 | 0.7450 | 0.9823 |  |  |  |  |  |  |
| Unknown (UNKN) | -0.4204 | 0.5526 | 0.1633 | -0.0567 | 0.0000 | -0.0037 | 0.0398 | 0.3728 | 0.5464 | 0.5848 | 10.1 | 0.0015 |
| (t-value) | -0.7616 | 0.9249 | 0.7579 | -0.7518 | -0.8653 | -0.1952 |  |  |  |  |  |  |
| (p-value) | 0.4485 | 0.3578 | 0.4507 | 0.4544 | 0.3894 | 0.8457 |  |  |  |  |  |  |

*Table 8: Manual Systems Main Results*

|  | Estimators | | | | | | | | Woolridge Individual Effects Test | | Woolridge Serial Correlation Test | |
|---|---|---|---|---|---|---|---|---|---|---|---|---|
|  | SOX 404 Effective Control | SOX 302 Effective Control | SOX 302 Mtl Weakness | SOX 302 Sig Deficiency | Audit Fees | Other Fees | $R^2$ | F | z | p-value | $\psi^2$ | p-value |
| Payment Card Fraud (CARD) | -0.9931 | 0.9877 | 0.0060 | -0.0239 | 0.0000 | 0.0023 | 0.1681 | 1.7736 | -0.8886 | 0.3742 | 4.3 | 0.0372 |
| (t-value) | -2.7572 | 2.5254 | 0.0425 | -0.4881 | -0.5314 | 0.2008 | | | | | | |
| (p-value) | 0.0072 | 0.0136 | 0.9662 | 0.6268 | 0.5966 | 0.8414 | | | | | | |
| Unintended disclosure (DISC) | 0.2688 | -0.1400 | 0.4254 | -0.0975 | 0.0000 | 0.0132 | 0.2340 | 2.6815 | 1.6445 | 0.1001 | 10.8 | 0.0010 |
| (t-value) | 0.3378 | -0.1620 | 1.3655 | -0.9024 | -0.0587 | 0.5287 | | | | | | |
| (p-value) | 0.7364 | 0.8717 | 0.1760 | 0.3696 | 0.9534 | 0.5985 | | | | | | |
| Hacking or malware (HACK) | 0.3213 | -0.2443 | -0.2500 | 0.0541 | 0.0000 | -0.0016 | 0.0461 | 0.4238 | 1.4766 | 0.1398 | 4.1 | 0.0436 |
| (t-value) | 0.3839 | -0.2688 | -0.7627 | 0.4764 | -0.0167 | -0.0626 | | | | | | |
| (p-value) | 0.7020 | 0.7888 | 0.4479 | 0.6351 | 0.9867 | 0.9502 | | | | | | |
| Insider (INSD). | -0.3278 | 0.5372 | -0.1096 | 0.1916 | 0.0000 | -0.0176 | 0.2255 | 2.5551 | 1.9611 | 0.0499 | 10.5 | 0.0012 |
| (t-value) | -0.4380 | 0.6610 | -0.3741 | 1.8846 | -1.4045 | -0.7520 | | | | | | |
| (p-value) | 0.6626 | 0.5105 | 0.7093 | 0.0632 | 0.1641 | 0.4543 | | | | | | |
| Physical loss (PHYS) | 0.1639 | 0.1037 | 0.3838 | 0.0578 | 0.0000 | 0.0517 | 0.3588 | 4.9115 | 0.5475 | 0.5840 | 124.3 | 0.0000 |
| (t-value) | 0.3017 | 0.1758 | 1.8044 | 0.7839 | -0.0238 | 3.0376 | | | | | | |
| (p-value) | 0.7636 | 0.8609 | 0.0750 | 0.4355 | 0.9811 | 0.0032 | | | | | | |
| Portable device (PORT) | 0.9591 | -0.7614 | 0.3866 | -0.1293 | 0.0000 | -0.0468 | 0.2031 | 2.2365 | 0.6680 | 0.5042 | 2.5 | 0.1118 |
| (t-value) | 0.9388 | -0.6863 | 0.9665 | -0.9319 | 1.6864 | -1.4618 | | | | | | |
| (p-value) | 0.3507 | 0.4945 | 0.3368 | 0.3542 | 0.0957 | 0.1478 | | | | | | |
| Stationary device (STAT) | 0.0497 | -0.0612 | -0.0158 | 0.0009 | 0.0000 | 0.0001 | 0.0015 | 0.0135 | 1.4697 | 0.1416 | 780210.0 | 0.0000 |
| (t-value) | 0.1838 | -0.2083 | -0.1491 | 0.0250 | 0.3263 | 0.0080 | | | | | | |
| (p-value) | 0.8546 | 0.8355 | 0.8819 | 0.9801 | 0.7451 | 0.9936 | | | | | | |
| Unknown (UNKN) | -0.4420 | 0.5782 | 0.1736 | -0.0538 | 0.0000 | -0.0011 | 0.0403 | 0.3690 | 0.5449 | 0.5859 | 10.1 | 0.0015 |
| (t-value) | -0.7860 | 0.9469 | 0.7884 | -0.7046 | -0.8802 | -0.0652 | | | | | | |
| (p-value) | 0.4342 | 0.3466 | 0.4328 | 0.4831 | 0.3814 | 0.9482 | | | | | | |

*Table 9: Automated Systems Main Results*

We tested a panel binomial models (logit and probit) using a General Linear Model approach, but achieved better fit using standard linear model panel regression. The reason that a binomial model may be preferred is to avoid estimator bias with smaller sample sizes, but in our research of $(breaches \sim SOX)_{B \cap C}$ the data sets are sufficiently large that there is no advantage to using probit or logit models, and linear model panel regression approaches will yield unbiased estimators. The appropriateness of the fixed effects linear model was tested using Hausman test, Chamberlain test for fixed effects, and Wooldridge's test for unobserved effects in panel models (Wooldridge 2010, Shang and Wooldridge 2016, Wooldridge 2016). We computed Wooldridge's test statistics for models with each type of breach dependent indicator variable. Test statistics for all the models indicate there are possible unobserved individual effects and thus we used fixed effects panel regression.

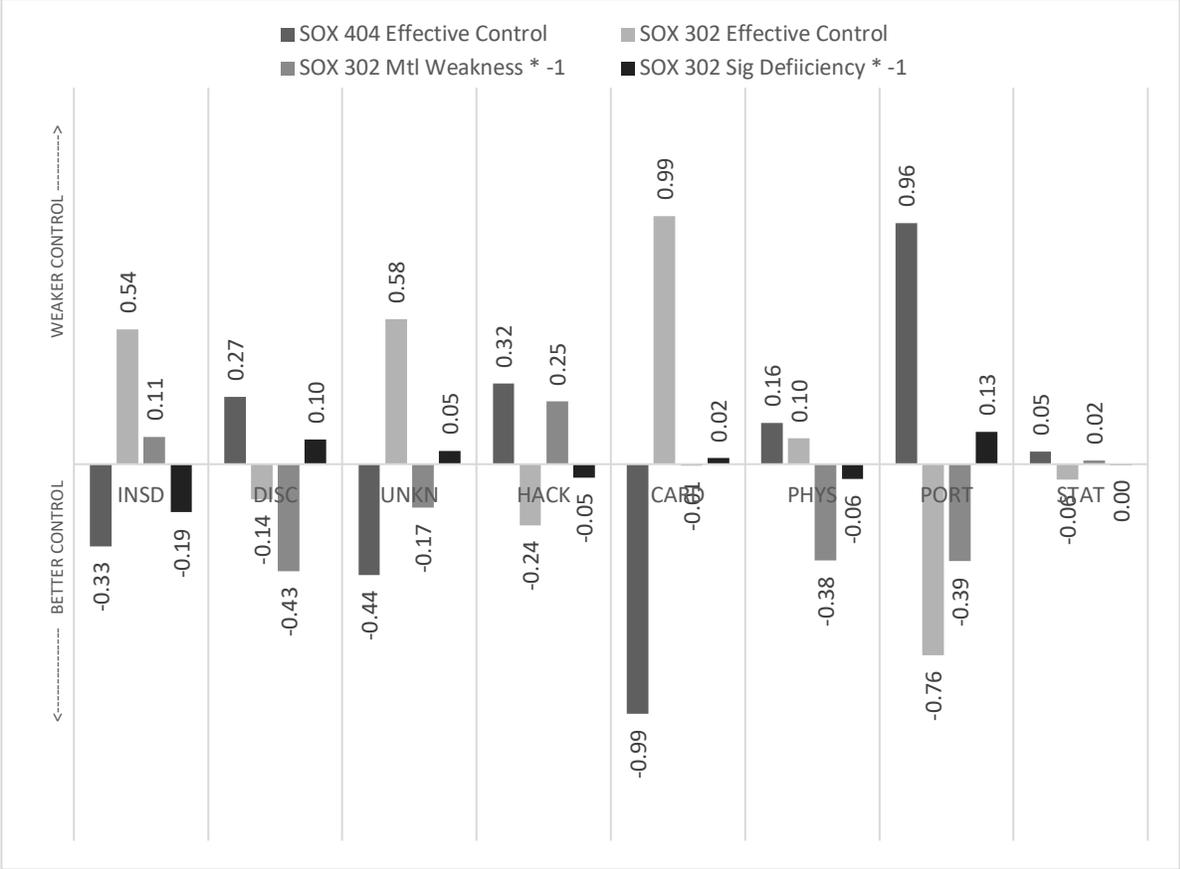

*Figure 2: Relative Effectiveness of SOX Assessments in Controlling Types of Security Breaches in Automated Systems*

Wooldridge's test for serial correlation in fixed effects statistics also suggested significant serial correlation in all the models, which seems reasonable given that security breaches often happen long before they are detected, and indeed may go uncorrected for some time after being detected. Note that an assessment of good internal control (SOX 302 and 404) is represented by an indicator value of "1". Assessments of material weaknesses and significant deficiencies are also represented by an indicator value of "1", but imply the contrary assessment that control is not adequate. Thus, in establishing the vertical axes in figures 2 and 3, the latter two estimators were multiplied by −1.

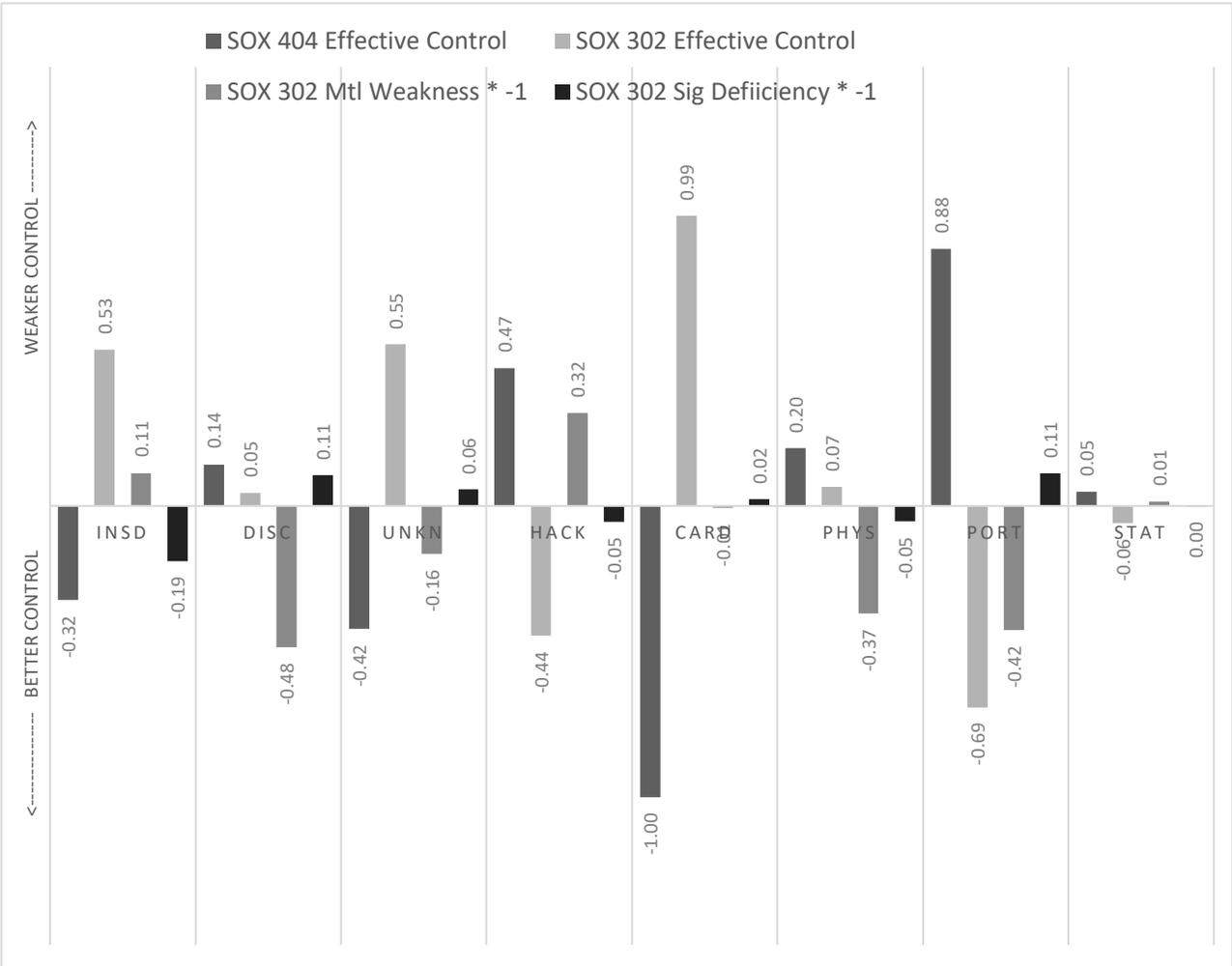

*Figure 3: Relative Effectiveness of SOX Assessments in Controlling Types of Security Breaches in Manual Systems*

Figures 2 and 3 displays the relative effectiveness of SOX assessments in controlling various types of security breaches in automated and manual systems respectively. Looking more closely at these effects, we can see that SOX auditing has significantly varying effectiveness in predicting the various classes of security breach, but the results were very similar for manual and automated systems. The model fit ($R^2$) was nearly zero for hacking and malware (HACK), stationary devices (STAT) and situations where the cause of the breach was unknown (UNKN), suggesting that current SOX audit practice both inside and outside of the firm provide little information useful in controlling these threats. The estimators for unintended disclosures (DISC) and physical losses (PHYS) were small, which also suggests that SOX audits provide little information useful for controlling these types of breaches. Where SOX provided significant predictive power in identifying future security breaches, management's (SOX 302) assessments were generally much less informative than those provided by SOX 404 audits. This concurs with what we discovered in our previous tests that did not include the breach information, and is consistent with prior research by (Benoit 2006, Bedard, Hoitash et al. 2009) who found that management is overconfident in assessing internal control effectiveness reflected in self-reporting under SOX302. SOX 404 adverse decisions on effectiveness of controls occurred in 100% of credit card (CARD) data breaches and around 33% of insider (INSD) breaches. But SOX 404 audits gave "effective" control decisions on 88% of situations where there was a control breach concerning a portable (PORT) device. This suggests that employees are subverting strict internal controls by using portable devices that can be carried outside the physical boundaries of the firm, and which can be used more freely than devices that are directly under the firm's control. Furthermore, the SOX auditors are not catching these subversions to internal control.

6. Structural Models

The panel regression results reported above provide useful insights into idiosyncrasies of SOX auditing and its effectiveness with respect to specific types of security breaches. Our $R^2$ fit statistics ranged from under 1% up to 36%, depending on the amount of information we have about specific types of breaches (breaches that are more common provided more data, and thus better model fit). These research datasets arose through natural experiments with biases that were discussed earlier, and we did not have the luxury of randomization as we would expect from laboratory studies. To see if we could improve on these fits, we imposed additional structure on our research models in a quest to infer the underlying processes generating the data and to potentially control potential biases arising from observational data. (Lucas 1976) argued that generic additive linear models such as those invoked in our previous section panel regressions lack stability and robustness. Empirical models are improved when constructs are policy-invariant i.e. structural i.e. not they would necessarily change whenever the competitive environment or policy was changed. (Lucas 1976) promoted a positive research program for how to do dynamic quantitative economics arguing that real-world decisions will diverge from the relationships observed in historical data. Policy-invariant structural models need to be constructed through analysis of the underlying dynamics of the construct relationships and behavior.

Invoking a theoretical model for the underlying security risks faced by firms solves two problems in our subsequent analysis:

1. Breach data is sparse so panel regression algorithms are more likely to terminate without a solution due to singular matrices or failure of the algorithm to converge. A theoretical distribution allows effective bootstrapping to resolve computation problems.
2. Breach data is significantly underreported and a theoretical model allows development of theoretically sound extrapolations to better estimate the true risk to firms of security breaches.

In this section, we developed structural models that aggregated panels and used empirical distributions of rate of breach occurrence using two alternative underlying generation models. This reduced our merged dataset to a dataset of individual firm "observations" that link breach occurrence timing distributions to SOX assessment timing distributions. The two structural models were a hazard model (time to breach / failure) and an occupancy model (breaches assigned to firm

### a) Hazard Structural Model of Breaches

We constructed the breach hazard dataset by reorganizing the full breach report dates (month-day-year) into intervals of days between a breach. The data was left-censored at January 1 2005 and right-censored at December 31 2016. A similar sort of structured model would not make sense on for the SOX dataset since SOX reports appear at regular annual intervals so instead the full SOX dataset and empirical "expected time to breach" parameter estimates for the analysis of our regression model ($empirical\ distribution\ of\ time\ to\ next\ breach \sim SOX)_{B \cap C}$ . We additionally had data on type of breach (e.g. a Hack, Inside Job etc.) but this was sparse making it problematic to make broader inferences in this way at any resolution less that the general reporting of breach.

We computed the times in days between breaches and fit a left-right censored (by these dates) dataset to a potential "time to next breach" distribution. Three distributions are commonly considered in this sort of failure analysis:

1. Exponential which assumes that breaches are independent of each other

2. Gamma offers a very general distribution that includes both exponential and Gaussian distributions as well as other candidates and

3. Gaussian offers a convenient distribution for additive processes with very well developed statistical algorithms.

Following our original assumption that hackers are agnostic that they do not favor breaching one firm over another and assuming they act independently of each other they do not tell many people about their illegal breaches we initially favored a working hypothesis that breach occurrences were distributed exponentially. Our maximum likelihood fit of "expected time to next breach" observations though strongly rejected the exponential distribution. This seems to imply that breaches are not stochastically independent events which makes sense if we consider an alternative hypothesis that hackers may cost-effectively attack many firms at once using the same exploits resulting in many successful hacks that are not stochastically independent of each other.

The Gamma distribution (which includes both exponential and Normal distributions as subclasses) in contrast provided a good fit giving very high values to the shape and scale parameters. As shape and scale increase the Gamma distribution converges to a Gaussian distribution and our empirical fit strongly supports expected time between breaches being Gaussian distributed. The mean time (in days) between breaches was estimated by fitting[2] the censored breach data (in the period 2005 through 2015) to a Gaussian distribution with mean and variance statistics computed for each listed firm and year for which SOX reports were generated.

Where a company had no breaches in the censored time interval we used a conservative estimate of a Gaussian distribution of time to next breach with mean = standard deviation = 3651 $days$ in the interval January 1 2005 to December 31 2014 reflecting a less than a 50% chance of a breach in this period.

*Table 10: Regression results from the Hazard Model of Breaches*

|  |  | (Intercept) | 302: Effective Control | 302: Material Weakness | 302:Signif Deficiency | 404: Effective Control | $R^2$ |
|---|---|---|---|---|---|---|---|
| Time (days) to Next Breach of Automated Systems | Estimate | 2047.774 | -58.689 | -1.634 | 18.657 | -175.264 | 0.66% |
|  | Std. Error | 149.608 | 112.124 | 151.709 | 36.502 | 119.805 |  |
|  | t-val | 13.688 | -0.523 | -0.011 | 0.511 | -1.463 |  |
| Time (days) to Next Breach of Manual Systems | Estimate | 2047.778 | -59.231 | -1.602 | 19.364 | -175.444 | 0.67% |
|  | Std. Error | 149.608 | 112.126 | 151.71 | 36.512 | 119.805 |  |
|  | t-val | 13.688 | -0.528 | -0.011 | 0.53 | -1.464 |  |

Table 10 summarizes the $(empirical\ distribution\ of\ time\ to\ next\ breach \sim SOX)_{B \cap C}$ regression using a hazard model that assumes a Gaussian distribution. Unfortunately, there is insufficient data on the individual types of breaches to fit hazard models, so we need to be content with estimation of the undifferentiated breach total. But $R^2$ is around $2/3$ for both manual and automated systems, which strongly supports the modeled relationships. Furthermore, we argue that the information provided from estimation of the hazard structural model is complementary to our other tests, as it estimates the change in expected time between breaches given management's or auditors' SOX attestation, and the results are consistent whether addressing manual or automated systems.

---

[2] Empirical fit was accomplished using the `fitdistcens` algorithm in the `fitdistrplus` package in R which computes maximum likelihood estimations of the distribution parameters using the Nelder-Mead method.

*Table 11: Change in Breach Frequency based on Hazard Model*

|  | Change in Number of Days between Breaches | % Change | Conclusion |
|---|---|---|---|
| *Expected days between breaches when SOX302 and SOX404 indicators = 0* | 2048 | 100.00% | Baseline |
| SOX302: Effective Control =1 | +59 | +2.88% | Less Frequent Breaches when Control is Effective |
| SOX302: Material Weakness =1 | +1.6 | +0.08% | No Effect |
| SOX302: Signif Deficiency =1 | -19 | -0.93% | Slightly More Frequent Breaches with Significant Deficiencies |
| SOX404: Effective Control =1 | +175 | +8.54% | Less Frequent Breaches when Control is Effective |

Table 11 provides policy recommendations and where management knows the expected cost of a breach, a specific value which may be matched to the cost of the SOX404 and SOX302 attestations. For example, if we assumed that the breach in question was a Home Depot or Target sized breach of say around $100 million occurring every 2048 days (5.6 years), then a 175 day increase represents an average $8.5 million annual savings by conducting the SOX404 audit. The hazard structural model is highly suggestive, but additional data at the level of specific type of breach will be needed before this model can be explored further.

### b) Occupancy Structural Model of Breaches

Our regression of $(empirical\ rank-frequency\ distribution\ of\ breaches \sim SOX)_B$ was extrapolated to the full set of firms *B* using a Bose-Einstein occupancy model. Applications of Bose-Einstein statistics in economics, psychology and finance have grown substantially over the past decade, e.g., in (Kürten and Kusmartsev 2011, Xu 2015) (Pascual-Leone 1970, Bouchaud and Mézard 2000, Weiss and Weiss 2003, Mezard and Montanari 2009) and (Amati and Van Rijsbergen 2002). They have been shown to provide accurate models of economic networks like the financial reporting and Internet hacker networks investigated here. They are effective in modeling competitive, non-equilibrium systems and can predict first-mover-advantage, Matthew Effects and winner-takes-all phenomena observed in competitive systems as phases of the underlying evolving networks (Bianconi and Barabási 2001).

Bose-Einstein model construction in this section proceeds as follows. Let there be $K$ listed companies on the NYSE+NASDAQ exchanges with $M_i$ security breaches in the $K$ companies' systems and $N_i$ wealth (dollars) that could be extracted from a breach of the systems of the $i^{th}$ company within an interval of time. This can be defined as an "occupancy" problem of a form extensively studied in physics where three main systems are defined: Fermi-Dirac Maxwell–Boltzmann and Bose-Einstein occupancy. Fermi-Dirac occupancy implies that a company's systems cannot be hacked or breached twice (i.e. lightening doesn't strike twice) in some time interval which is clearly not realistic. Maxwell–Boltzmann occupancy is a theoretical concept at the core of the Gibbs Paradox and does not make sense in the current research context. Bose-Einstein occupancy in contrast is found in many real-world phenomena and is a basic assumption that we make in this analysis of corporate security breaches. Bose-Einstein occupancy implies that those who commit security breaches are agnostic – they don't care what firm they hack; they are attracted instead by any chance for gain, havoc or other hacking goal.

Let $N_i$ be large $M_i$ random and make the very weak assumption of convergence to a limiting distribution $\frac{M_i}{N_i} \to F(x) \sim Cx^\gamma$ as $x \to 0$ $\gamma > 0$. (Hill 1974) building on earlier work from (Simon 1955, Mandelbrot 1960, Mandelbrot 1965) and (Hill 1970) showed that if the rank-frequency distribution of number of security breaches $L^{(r)}$ to the systems of the $r^{th}$ company follows a Bose-Einstein allocation of $N_i$ wealth (dollars) extracted from the company through $M_i$ hacking attempts then $L^{(r)} \propto r^{-(1+\alpha)}$ for $1 + \alpha = \frac{1}{\gamma}$. This is called a Zipf distribution, and has mass function (and thus frequency) $f_{1+\alpha}(r) = r^{-(1+\alpha)} / \zeta(r)$ where $\zeta(r)$ is Reimann's zeta and $\alpha \in [0, \infty]$. In our dataset $K = 3398$ NYSE+NASDAQ listed firms and of the 388 breaches in the data $\sum_{\forall i} M_i = 213$ occurred in 184 of the NYSE+NASDAQ listed firms ($B \cap C$) in the time interval from 2005 to 2015.

*Table 12: Estimates of α for frequencies* $f_{1+\alpha}(r) \propto r^{-(1+\alpha)}$

|  | CARD | DISC | HACK | INSD | PHYS | PORT | STAT | UNKN |
|---|---|---|---|---|---|---|---|---|
| $Estimate\ of\ \alpha$ | 2.0008 | 3.9656 | 4.4153 | 3.5513 | 2.4404 | 4.3913 | 1.9381 | 2.1775 |
| $Regression\ Estimator$ | -1.0008 | -2.9656 | -3.4153 | -2.5513 | -1.4404 | -3.3913 | -0.9381 | -1.1775 |
| $Std.Error$ | 0.0664 | 0.1084 | 0.1271 | 0.0990 | 0.0778 | 0.1255 | 0.0645 | 0.0713 |
| $t-value$ | -15.0800 | -27.3500 | -26.8600 | -25.7640 | -18.5200 | -27.0200 | -14.5500 | -16.5040 |
| $R^2$ | 37.06% | 65.96% | 65.15% | 63.23% | 47.06% | 65.41% | 35.43% | 41.37% |
| $F-stat$ | 227 | 748 | 722 | 664 | 343 | 730 | 212 | 272 |
| $E\ (days\ between\ breach)$ $right-tail\ only$ | 13,203 | 4,773 | 2,289 | 6,919 | 11,825 | 2,434 | 13,366 | 12,697 |

Table 12 shows the results of fitting our breach dataset to the occupancy model. All regressions were significant at the .0001 significance level, despite three limitations in our data:

1. Industry sources suggest that security breaches are substantially underreported by factors ranging from 3 to 30 (Menn 2012, Symantic 2016).
2. Breach counts are integer and thus the estimated continuous curve is an approximation.
3. All but 184 (4.6%) of the 3398 listed firms reported zero breaches in the research period; these clearly represent the range where $L^{(r)} < 1$ since breach counts are integers.

The absence of reported breaches in the right tail of our rank-frequency ordering does not imply that these firms face no threat from security breaches. Rather breach probabilities in the right tail reflect a pervasive low level background threat facing all firms equally (since firms with no breaches cannot be ranked) and these should be treated similarly to so-called 5000 year errors (Adams 1980) in software testing. From our perspective, the firms in the right tail are unranked and identical and thus face a uniform background probability of breach that is very low but not zero.

The empirical coefficient $\alpha$ has various explanations in the literature depending on whose model you are using; for example (Mandelbrot 1960, Mandelbrot 1965) contends that these represent a sort of systems entropy and can be modeled using fractals. It is beyond the scope of this research to further embellish such narratives and we simply assume that the value of $\alpha$ describes some artifact or metric of industry wide security.

Zipfian rank-frequency distributions are fat-tailed; their probability density function is proportional to $r^{-a} \to 0$ for large $r$ and the cumulative tail distribution does not converge. In our case the tail is right-bounded at $r = 3398$. The breach probability for a single firm in the right tail is

$$\frac{\sum_{r=185}^{3398} r^{-(1+\alpha)}}{\sum_{r=1}^{3398} r^{-(1+\alpha)}}$$

and the mean days between breaches will be

$$\frac{\sum_{r=1}^{3398} r^{-(1+\alpha)} \times 10 \text{ years} \times 365.25 \text{ days}/\text{year}}{\sum_{r=185}^{3398} r^{-(1+\alpha)}}$$

The mean time to failure ranges from 6 to 36 years for individual breach types, and over 5 months for all types combined. This is somewhat more frequent than the 5.6 years expected between breaches by the hazard structural model, but with more extensive data collection, a more robust estimate could be obtained, and we consider this estimate to be realistic for the occupancy structural model. Such failures where they do occur in the software field are apocryphally called 5000 errors and have been studied in (Boehm, Clark et al. 1995, Westland 2000, Westland 2002, Westland 2004) .

*Table 13:* Occupancy Structural Model for **Automated** Systems

| | | SOX 302 | | SOX 404 | | Fees | | | | |
|---|---|---|---|---|---|---|---|---|---|---|
| *Estimators* | IC Effective | Mat Weakness | Sig Defic | IC Effective | Audi | Non-Audit | Tax | Related | Other | $R^2$ and $F-stat$ |
| CARD | 8.66E-14 | 9.52E-14 | -1.41E-14 | -4.14E-14 | 1.21E-19 | -2.09E-17 | 1.95E-17 | 2.19E-17 | 2.05E-17 | 0.001653 |
| s.e. | 8.84E-14 | 5.62E-14 | 2.59E-14 | 4.41E-14 | 8.21E-19 | 3.42E-16 | 3.42E-16 | 3.42E-16 | 3.42E-16 | 0.5402 |
| t-value | 0.979 | 1.693 | -0.544 | -0.938 | 0.148 | -0.061 | 0.057 | 0.064 | 0.06 | |
| p-value | 0.3276 | 0.0905 | 0.5862 | 0.3482 | 0.8825 | 0.9513 | 0.9546 | 0.9491 | 0.9522 | |
| DISC | 2.18E-20 | 2.46E-20 | -3.46E-21 | -1.05E-20 | 3.15E-26 | -3.92E-24 | 3.53E-24 | 4.18E-24 | 3.82E-24 | 0.001601 |
| s.e. | 2.21E-20 | 1.41E-20 | 6.49E-21 | 1.10E-20 | 2.06E-25 | 8.58E-23 | 8.58E-23 | 8.57E-23 | 8.58E-23 | 0.5234 |
| t-value | 0.983 | 1.744 | -0.533 | -0.951 | 0.153 | -0.046 | 0.041 | 0.049 | 0.044 | |
| p-value | 0.3258 | 0.0812 | 0.5943 | 0.3415 | 0.8784 | 0.9636 | 0.9672 | 0.9611 | 0.9645 | |
| HACK | 6.53E-22 | 7.39E-22 | -1.04E-22 | -3.15E-22 | 9.59E-28 | -1.15E-25 | 1.03E-25 | 1.23E-25 | 1.12E-25 | 0.001599 |
| s.e. | 6.65E-22 | 4.23E-22 | 1.95E-22 | 3.31E-22 | 6.17E-27 | 2.57E-24 | 2.58E-24 | 2.57E-24 | 2.57E-24 | 0.5225 |
| t-value | 0.982 | 1.747 | -0.531 | -0.951 | 0.155 | -0.045 | 0.04 | 0.048 | 0.043 | |
| p-value | 0.3261 | 0.0807 | 0.5952 | 0.3415 | 0.8765 | 0.9645 | 0.9681 | 0.962 | 0.9654 | |
| INSD | 5.47E-19 | 6.15E-19 | -8.70E-20 | -2.64E-19 | 7.75E-25 | -1.02E-22 | 9.19E-23 | 1.08E-22 | 9.89E-23 | 0.001605 |
| s.e. | 5.56E-19 | 3.54E-19 | 1.63E-19 | 2.77E-19 | 5.16E-24 | 2.15E-21 | 2.15E-21 | 2.15E-21 | 2.15E-21 | 0.5245 |
| t-value | 0.984 | 1.74 | -0.534 | -0.951 | 0.15 | -0.047 | 0.043 | 0.05 | 0.046 | |
| p-value | 0.3254 | 0.0819 | 0.5933 | 0.3416 | 0.8806 | 0.9624 | 0.966 | 0.96 | 0.9634 | |
| PHYS | 2.96E-15 | 3.29E-15 | -4.77E-16 | -1.42E-15 | 4.02E-21 | -6.42E-19 | 5.92E-19 | 6.76E-19 | 6.27E-19 | 0.001627 |
| s.e. | 3.01E-15 | 1.92E-15 | 8.83E-16 | 1.50E-15 | 2.80E-20 | 1.17E-17 | 1.17E-17 | 1.17E-17 | 1.17E-17 | 0.532 |
| t-value | 0.983 | 1.715 | -0.541 | -0.946 | 0.144 | -0.055 | 0.051 | 0.058 | 0.054 | |
| p-value | 0.3255 | 0.0864 | 0.5889 | 0.3442 | 0.8858 | 0.9561 | 0.9595 | 0.9538 | 0.9571 | |
| PORT | 7.87E-22 | 8.91E-22 | -1.25E-22 | -3.80E-22 | 1.16E-27 | -1.39E-25 | 1.24E-25 | 1.48E-25 | 1.35E-25 | 0.001599 |
| s.e. | 8.02E-22 | 5.10E-22 | 2.35E-22 | 4.00E-22 | 7.44E-27 | 3.11E-24 | 3.11E-24 | 3.10E-24 | 3.10E-24 | 0.5226 |
| t-value | 0.982 | 1.747 | -0.531 | -0.951 | 0.155 | -0.045 | 0.04 | 0.048 | 0.043 | |
| p-value | 0.3261 | 0.0807 | 0.5951 | 0.3415 | 0.8766 | 0.9644 | 0.9681 | 0.962 | 0.9654 | |
| STAT | 1.40E-13 | 1.54E-13 | -2.28E-14 | -6.67E-14 | 1.98E-19 | -3.44E-17 | 3.21E-17 | 3.59E-17 | 3.38E-17 | |
| s.e. | 1.43E-13 | 9.09E-14 | 4.19E-14 | 7.13E-14 | 1.33E-18 | 5.54E-16 | 5.54E-16 | 5.54E-16 | 5.54E-16 | 0.001658 |
| t-value | 0.978 | 1.689 | -0.545 | -0.936 | 0.149 | -0.062 | 0.058 | 0.065 | 0.061 | 0.542 |
| p-value | 0.3281 | 0.0913 | 0.5858 | 0.3491 | 0.8813 | 0.9504 | 0.9537 | 0.9482 | 0.9514 | |
| UNKN | 2.24E-14 | 2.47E-14 | -3.62E-15 | -1.07E-14 | 3.07E-20 | -5.15E-18 | 4.78E-18 | 5.39E-18 | 5.04E-18 | 0.00164 |
| s.e. | 2.28E-14 | 1.45E-14 | 6.67E-15 | 1.14E-14 | 2.12E-19 | 8.82E-17 | 8.82E-17 | 8.82E-17 | 8.82E-17 | 0.5362 |
| t-value | 0.981 | 1.703 | -0.543 | -0.942 | 0.145 | -0.058 | 0.054 | 0.061 | 0.057 | |
| p-value | 0.3264 | 0.0886 | 0.5874 | 0.3462 | 0.8848 | 0.9535 | 0.9568 | 0.9512 | 0.9545 | |
| Total Breach | 2.52E-13 | 2.77E-13 | -4.10E-14 | -1.20E-13 | 3.54E-19 | -6.11E-17 | 5.70E-17 | 6.39E-17 | 5.99E-17 | 0.001654 |
| s.e. | 2.57E-13 | 1.64E-13 | 7.53E-14 | 1.28E-13 | 2.39E-18 | 9.96E-16 | 9.96E-16 | 9.96E-16 | 9.96E-16 | 0.5407 |
| t-value | 0.979 | 1.692 | -0.545 | -0.938 | 0.148 | -0.061 | 0.057 | 0.064 | 0.06 | |
| p-value | 0.3277 | 0.0907 | 0.5861 | 0.3485 | 0.8821 | 0.9511 | 0.9544 | 0.9489 | 0.952 | |

*Table 14:* Occupancy Structural Model for **Manual** Systems

| | SOX 302 | | | SOX 404 | | Fees | | | | |
|---:|---|---|---|---|---|---|---|---|---|---|
| Estimators | IC Effective | Mat Weakness | Sig Defic | IC Effective | Audit | Non-Audit | Tax | Related | Other | $R^2$ and $F-stat$ |
| CARD | 8.25E-14 | 8.47E-14 | -1.05E-14 | -2.63E-14 | 2.06E-19 | -1.77E-17 | 1.62E-17 | 1.84E-17 | 1.66E-17 | 0.001414 |
| s.e. | 7.00E-14 | 5.15E-14 | 2.31E-14 | 2.32E-14 | 7.65E-19 | 3.20E-16 | 3.20E-16 | 3.20E-16 | 3.20E-16 | 0.5272 |
| t-value | 1.178 | 1.644 | -0.456 | -1.133 | 0.269 | -0.055 | 0.051 | 0.058 | 0.052 | |
| p-value | 0.239 | 0.1 | 0.649 | 0.257 | 0.788 | 0.956 | 0.96 | 0.954 | 0.959 | |
| DISC | 2.06E-20 | 2.17E-20 | -2.56E-21 | -6.63E-21 | 5.19E-26 | -3.28E-24 | 2.89E-24 | 3.50E-24 | 3.00E-24 | 0.001357 |
| s.e. | 1.75E-20 | 1.29E-20 | 5.78E-21 | 5.81E-21 | 1.92E-25 | 8.02E-23 | 8.02E-23 | 8.02E-23 | 8.02E-23 | 0.5059 |
| t-value | 1.176 | 1.683 | -0.443 | -1.141 | 0.271 | -0.041 | 0.036 | 0.044 | 0.037 | |
| p-value | 0.2396 | 0.0925 | 0.6578 | 0.254 | 0.7867 | 0.9674 | 0.9712 | 0.9652 | 0.9702 | |
| HACK | 6.18E-22 | 6.52E-22 | -7.67E-23 | -1.99E-22 | 1.57E-27 | -9.60E-26 | 8.42E-26 | 1.03E-25 | 8.74E-26 | 0.001354 |
| s.e. | 5.26E-22 | 3.87E-22 | 1.74E-22 | 1.75E-22 | 5.75E-27 | 2.41E-24 | 2.41E-24 | 2.41E-24 | 2.41E-24 | 0.5048 |
| t-value | 1.175 | 1.685 | -0.442 | -1.141 | 0.273 | -0.04 | 0.035 | 0.043 | 0.036 | |
| p-value | 0.2401 | 0.0921 | 0.6587 | 0.2541 | 0.785 | 0.9682 | 0.9721 | 0.966 | 0.9711 | |
| INSD | 5.18E-19 | 5.44E-19 | -6.45E-20 | -1.67E-19 | 1.29E-24 | -8.52E-23 | 7.55E-23 | 9.06E-23 | 7.79E-23 | 0.001361 |
| s.e. | 4.40E-19 | 3.24E-19 | 1.45E-19 | 1.46E-19 | 4.81E-24 | 2.01E-21 | 2.01E-21 | 2.01E-21 | 2.01E-21 | 0.5075 |
| t-value | 1.177 | 1.68 | -0.445 | -1.141 | 0.268 | -0.042 | 0.037 | 0.045 | 0.039 | |
| p-value | 0.2391 | 0.0931 | 0.6566 | 0.2539 | 0.7886 | 0.9663 | 0.9701 | 0.9641 | 0.9691 | |
| PHYS | 2.82E-15 | 2.92E-15 | -3.55E-16 | -9.01E-16 | 6.87E-21 | -5.41E-19 | 4.91E-19 | 5.69E-19 | 5.03E-19 | 0.001387 |
| s.e. | 2.39E-15 | 1.76E-15 | 7.87E-16 | 7.91E-16 | 2.61E-20 | 1.09E-17 | 1.09E-17 | 1.09E-17 | 1.09E-17 | 0.5172 |
| t-value | 1.18 | 1.661 | -0.451 | -1.139 | 0.263 | -0.05 | 0.045 | 0.052 | 0.046 | |
| p-value | 0.238 | 0.0968 | 0.6518 | 0.2549 | 0.7922 | 0.9605 | 0.9641 | 0.9584 | 0.9633 | |
| PORT | 7.46E-22 | 7.87E-22 | -9.25E-23 | -2.40E-22 | 1.89E-27 | -1.16E-25 | 1.02E-25 | 1.24E-25 | 1.06E-25 | 0.001354 |
| s.e. | 6.35E-22 | 4.67E-22 | 2.09E-22 | 2.11E-22 | 6.94E-27 | 2.91E-24 | 2.91E-24 | 2.90E-24 | 2.91E-24 | 0.5048 |
| t-value | 1.175 | 1.685 | -0.442 | -1.141 | 0.273 | -0.04 | 0.035 | 0.043 | 0.036 | |
| p-value | 0.2401 | 0.0921 | 0.6586 | 0.254 | 0.7851 | 0.9682 | 0.9721 | 0.966 | 0.971 | |
| STAT | 1.33E-13 | 1.37E-13 | -1.70E-14 | -4.25E-14 | 3.36E-19 | -2.91E-17 | 2.68E-17 | 3.03E-17 | 2.75E-17 | 0.00142 |
| s.e. | 1.13E-13 | 8.33E-14 | 3.73E-14 | 3.75E-14 | 1.24E-18 | 5.18E-16 | 5.18E-16 | 5.18E-16 | 5.18E-16 | 0.5293 |
| t-value | 1.178 | 1.64 | -0.456 | -1.132 | 0.271 | -0.056 | 0.052 | 0.058 | 0.053 | |
| p-value | 0.239 | 0.101 | 0.648 | 0.258 | 0.786 | 0.955 | 0.959 | 0.953 | 0.958 | |
| UNKN | 2.13E-14 | 2.19E-14 | -2.70E-15 | -6.79E-15 | 5.24E-20 | -4.34E-18 | 3.97E-18 | 4.55E-18 | 4.07E-18 | 0.001401 |
| s.e. | 1.80E-14 | 1.33E-14 | 5.95E-15 | 5.98E-15 | 1.97E-19 | 8.25E-17 | 8.25E-17 | 8.25E-17 | 8.25E-17 | 0.5224 |
| t-value | 1.179 | 1.652 | -0.454 | -1.136 | 0.266 | -0.053 | 0.048 | 0.055 | 0.049 | |
| p-value | 0.2383 | 0.0987 | 0.6501 | 0.2561 | 0.7906 | 0.9581 | 0.9616 | 0.9561 | 0.9607 | |
| Total Breach | 2.40E-13 | 2.46E-13 | -3.06E-14 | -7.65E-14 | 6.01E-19 | -5.16E-17 | 4.75E-17 | 5.38E-17 | 4.86E-17 | 0.001416 |
| s.e. | 2.04E-13 | 1.50E-13 | 6.72E-14 | 6.75E-14 | 2.23E-18 | 9.32E-16 | 9.32E-16 | 9.32E-16 | 9.32E-16 | 0.5278 |
| t-value | 1.178 | 1.643 | -0.456 | -1.132 | 0.27 | -0.055 | 0.051 | 0.058 | 0.052 | |
| p-value | 0.239 | 0.1 | 0.649 | 0.258 | 0.787 | 0.956 | 0.959 | 0.954 | 0.958 | |

The probability of a breach at any given time is small, with a large enough population of firms one can expect many failures to show up every year across the full population of firms and industry wide risk of breach is thus substantial. Even more so if you subscribe to the estimate that only 3-30% of breaches are reported, and thus in our dataset.

The $R^2$ and $F-statistics$ are only weakly significant or insignificant in all the regressions. The occupancy structural model does not contribute the hoped for new information on the 95.4% of the listed firms that did not publicly report breaches. Occupancy structural models may be useful in future estimation of breach probabilities for the full set of listed and SOX attested firms, but we will need substantially information about breaches that occur to successfully use such models for inference.

## 7. Causal Direction (counterfactuals)

The prior section tested whether managerial and audit exceptions in the SOX 302 and 404 attestations influence the probability of security breaches. Support for $H_1$ and rejection of $H_2$ and $H_3$ prove that adverse published opinions from SOX audits and lower audit fees are correlated with increased occurrence of security breaches. Our working hypothesis has been that the extensiveness of the audit opinion provides detailed guides to changes in a firm's internal control systems that will lower their probability of incurring a security breach in the future. To validate this causal chain:

$$adverse\ audit\ opinions \rightarrow improvements\ in\ IC \rightarrow lower\ risk\ of\ breach$$

To validate this we construct the counterfactual hypothesis to be tested (disproved):

> $H_4$: There will be higher cumulative abnormal SOX metrics in years following a security breach that indicate that the security breach has motivated auditors and management to render adverse opinions on the firm's internal controls.

To establish '*occurrence*' causality we tested the counterfactual $\{SOX|breach\ occurence\}$ i.e. whether the occurrence of a security breaches affect SOX 302 and 404 metrics using the Welch Two Sample t-test (table 3). To establish '*temporal*' causality we tested the counterfactual $\{SOX|breach\ date\}$ on the subset of 213 breaches in the largest registered firms following the (MacKinlay 1997) flow of analysis of event studies. We had 213 breach "events" recorded for NYSE and NASDAQ companies $\{b_{tic_1\ s_1} \ldots b_{tic_{213}\ s_{213}}\}$ at times $\{s_1 \ldots s_{213}\}$ and companies with a subset of ticker symbols $\{tic_{j_1} \ldots tic_{j_{213}}\} \in \{tic_1 \ldots tic_N\}$. For each breach event $b_{tic_{j_i}\ s_i}$ we define two time period durations $\{d_{pre-s_i}\ d_{post-s_i}\}$ For any particular SOX 302 or 404 metric (i.e. one of the columns in the dataset) we will have a market expectation (average) value for any SOX metric $E_{\forall it}(z_{imt})$ and abnormal return for firm $i$ time $t$ and metric $m$ $AR_{imt \in d_{period}} = z_{imt \in d_{period}} - E_{\forall it}(z_{imt})$ with $CAR_{im}^{d_{period}} = \sum_{\forall t \in d_{period}} AR_{imt}$. Security breach events were chosen from the period 2007 – 2014 which allowed two years on either side to include the full data series from our research dataset.

Appraisal of the event's impact requires a measure of abnormal return. The abnormal return is the actual ex post return of the asset over the event window minus the normal return of the asset over the event window. We used a constant mean return model and the normal return is defined as the expected return conditioned on the event. For firm $i$ and event date $t$ the abnormal return is

$$AR_{it} = R_{it} - E(R_{it}|X_t) \qquad (6)$$

where $AR_{it}$ $R_{it}$ and $E(R_{it}|X_t)$ are the abnormal actual and normal returns respectively for time period $t$. $X_t$ is the conditioning information from the normal return model and a return is computed from the dataset values (designated here as $P$ for consistency with event study notation).

$$R_{it} = \frac{P_{it} - P_{it-1}}{P_{it-1}} \qquad (7)$$

In the current research our data is categorical $P_{it} \in \{01\}$ and without loss of information we can set $R_{it} \cong P_{it}$ since all the $P_{it} \in \{01\}$. The abnormal return analogue for such data is

$$AR_{it} = P_{it} - E_{\forall it}(P_{it}) \qquad (8)$$

The cumulative abnormal return is:

$$CAR_i^{period} = \sum_{\forall t \in period} AR_{it} \qquad (9)$$

where the *period* is either the pre- or post- event 2-year window.

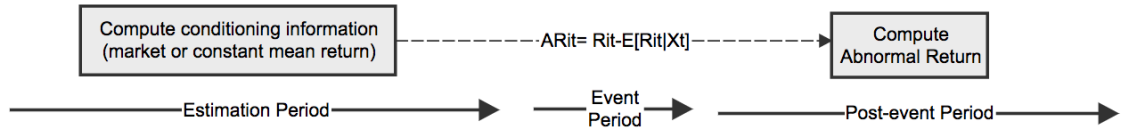

*Figure 4: Conceptual Schematic of Abnormal Return Calculations*

Since this research involves an annual time series the 120-day period typically used in stock price based event studies is too short. Rather we will assume the pre- and post- event windows are two years wide; and we will assume that there is no conditioning window around the event itself (as may be the case in stock price event studies). Security breach events were chosen from the period 2007 – 2014 which allowed two years on either side to include the full data series from our research dataset.

*Table 15: Occurrence Counterfactual: Welch Two Sample t-test comparing firms with and without breach in the estimation period for key SOX metrics*

|  | Welch t-stat | Welch-Satterthwaite DF | p-value | 95% Confidence Interval | |
|---|---|---|---|---|---|
| IS_EFFECTIVE | -0.602 | 336.540 | 0.548 | -0.026 | 0.014 |
| MATERIAL_WEAKNESS | 1.056 | 336.720 | 0.292 | -0.009 | 0.029 |
| SIG_DEFICIENCY | -1.285 | 336.060 | 0.200 | -0.053 | 0.011 |
| AUDITOR_AGREES | -0.089 | 89.372 | 0.929 | -0.099 | 0.090 |
| COMBINED_IC_OP | -0.227 | 122.490 | 0.821 | -0.069 | 0.055 |
| IC_IS_EFFECTIVE | 0.308 | 366.450 | 0.758 | -0.018 | 0.024 |
| AUDIT_FEES | -6.400 | 122.060 | 0.000 | -13851669 | -7306656 |

*Table 16: Temporal Counterfactual: Welch Two Sample t-test of CAR before and after breach event for key SOX metrics*

|  | Welch t-stat | Welch-Satterthwaite DF | p-value | 95% Confidence Interval | |
|---|---|---|---|---|---|
| IS_EFFECTIVE | -0.720 | 230.683 | 0.212 | -0.011 | 0.004 |
| MATERIAL_WEAKNESS | 1.095 | 71.641 | 0.298 | -0.003 | 0.029 |
| SIG_DEFICIENCY | -0.101 | 652.146 | 0.328 | -0.103 | 0.021 |
| AUDITOR_AGREES | -0.007 | 113.927 | 1.247 | -0.174 | 0.079 |
| COMBINED_IC_OP | -0.129 | 190.568 | 0.110 | -0.018 | 0.078 |
| IC_IS_EFFECTIVE | 0.329 | 394.594 | 1.142 | -0.019 | 0.038 |
| AUDIT_FEES | -12.445 | 187.358 | 0.000 | -13692930.748 | -66609.057 |

Tables 15 and 16 provide results of Welch Two Sample t-test's, which is the most suitable test for comparisons of the averages of two independent groups, extracted from two populations where variance is unknown and the sample variances are not homogeneous. Unfortunately, all the firm and auditors' SOX attestations p-values are quite high, and we are unable to differentiate the means of the two groups. The direction of causality is equivocal in all cases except for those related to audit fees. Figure 5 shows the consistency in the mean value of audit and non-audit fees charged to clients over time. Audit fees (black) rose sharply from 2003 to about 2007 because of SOX compliance work, but dropped back down and dipped after the 2008 recession. This created a situation where the cumulative average returns pre- and post-breach were different for the audit fees.

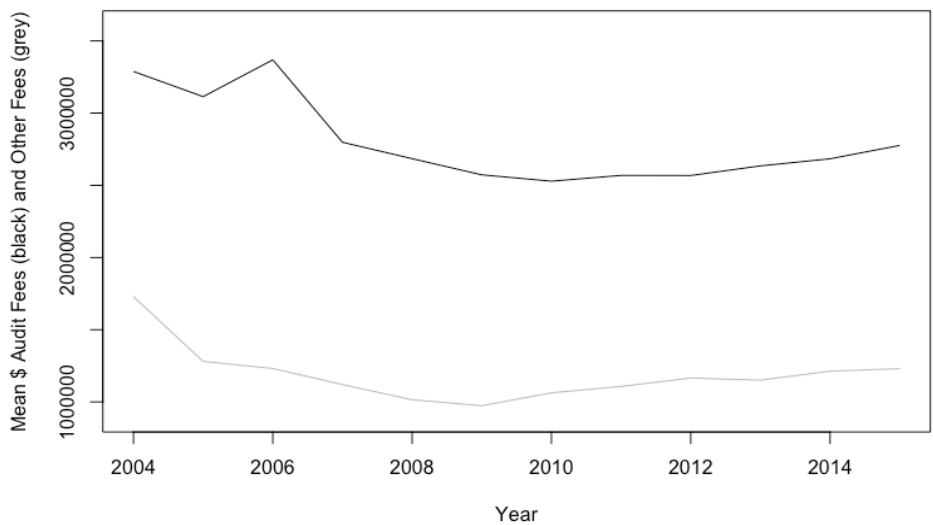

*Figure 5: Trends in Audit Fees (black) vs. Other Fees (grey) across firms in the dataset B ∩ C*

## 8. Conclusions and Discussion

How effective is corporate investment in SOX compliance in identifying and eliminating the threat from breaches? To that end we investigated in this paper breaches of corporate systems to determine whether SOX assessments are information bearing with respect to the firm's vulnerability to and history of security breaches. SOX focuses on a firm's internal control and its potential to suffer systems intrusion from external actors, which can lead to materially significant losses and misstatements.

We investigated $H_1$ whether the costly and controversial SOX 404 attestation work of external auditors provided information not in management attestations, published in SOX 302 reporting. We found for reports of both manual and automated systems that firms on average attest that internal controls are effective six times as often in SOX 302 reports as in SOX 404 reports. There was no evidence that any of the fees charged to the firm influenced the auditor's opinion – we validated the independence of SOX auditors. In both manual and automated systems, the firm's self-reported "material weaknesses" were associated with around 80% of SOX 404 negative control assessments, while self-reported "significant deficiencies" appeared in only around 1-2% of SOX 404 negative control assessments. This suggests that management's self-reported "material weaknesses" were a significant indicator of control problems identified by auditors, while significant deficiencies were not. Our findings strongly support the hypothesis $H_1$ that SOX 404 testing adds new information to self-reported SOX 302 attestations, and suggest that SOX 404 audits, though influenced by management's self-reported "material weaknesses," are additionally providing investors with information not otherwise available to stakeholders.

Hypothesis $H_2$ tested whether external auditors in their SOX 404 procedures were investigating the same systems control weaknesses, and finding the same types of problems, as management and internal auditors in their SOX 302 procedures. Complementing hypothesis $H_1$ we wanted to not only know whether external auditors were providing additional information, but also whether they were searching for the same set of control weaknesses as management's SOX 302 attestations. In the reports of both manual and automated systems, the estimators indicate that firms on average attest that internal controls are effective one-third to one-half as often in SOX 302 reports as in SOX 404 reports. Overall, firms seem to be substantially more vigilant and suspicious of internal control weaknesses than SOX 404 auditors. It is likely that management and internal auditors have more information about control weaknesses, since they audit the firm year-round, as opposed to the annual engagements of SOX404 teams. Magnitude changes on the SOX 302 "control is effective" and "material weakness" estimators as well as the sign reversals in "material weakness" estimates lead us to reject $H_2$. The sign reversal from panel regression results of the "material weakness" estimates suggests that the absence of "material weaknesses" self-reported by management in the SOX 302 reports was associated with significantly higher incidence of negative SOX 404 control assessments in general. This is consistent with prior research findings in (Benoit 2006, Bedard,

Hoitash et al. 2009) that management is overconfident in assessing internal control effectiveness reflected in self-reporting under SOX302 where 90% of companies with ineffective Section 404 controls still self-reported effective 302 controls in the same period end that an adverse Section 404 was reported. We rejected $H_2$ and found that management and auditors do not general agree on the underlying internal control situation at any time; instead the SOX 404 team was likely to discover material weaknesses and "educate" management and internal audit teams about the importance of these control weaknesses.

Our tests to this point have shown that the occurrence of breaches is only weakly affected by size, and that the strength of internal controls is likely to be a significant factor in the occurrence of a breach at a firm in a period. We also concluded that SOX 404 auditors maintain their independence consistent with Generally Accepted Auditing Standard (GAAS). With the investigation of potentially confounding factors of size and auditor independence effectively completed, we can move on to testing our main hypothesis $H_3$. We found that SOX 404 audits provided significantly more predictive power in identifying future security breaches than management's (SOX 302) assessments which is consistent with prior research findings in (Benoit 2006, Bedard, Hoitash et al. 2009, Rice and Weber 2012, Rice, Weber et al. 2014). SOX attestations were poor at identifying control weaknesses that might lead to unintended disclosures (DISC), physical losses (PHYS), hacking and malware (HACK), stationary devices (STAT) and situations where the cause of the breach was unknown (UNKN), suggesting that current SOX audit practice both inside and outside of the firm provide little information useful in controlling these threats. SOX 404 adverse decisions on effectiveness of controls occurred in 100% of credit card (CARD) data breaches and around 33% of insider (INSD) breaches. But SOX 404 audits gave "effective" control decisions on 88% of situations where there was a control breach concerning a portable (PORT) device, suggesting that employees are subverting particularly strict internal controls by using portable devices that can be carried outside the physical boundaries of the firm, and which can be used more freely than devices that are directly under the firm's control. And furthermore, the SOX auditors are not catching these subversions to internal control.

Hypothesis $H_4$ tested whether structural models as recommended in Lucas' critique (Lucas 1976) could extend our inferences from the relatively limited breach data by positing an underlying generation model. We constructed two models: (1) a hazard structural model and (2) an occupancy structural model. The hazard model provided specific management policy recommendations, especially where management knows the expected cost of a breach. We found an expected 2.88% reduction in breaches when SOX 302 controls are effective; management "material weakness' attestations provided no information in this structural model, whereas there would be around a 1% increase in breach occurrence when there are significant deficiencies. Finally, SOX 404 attestations were the most informative, and a negative SOX 404 attestation is projected to increase the frequency of breaches by around 8.5%. The hazard structural model is highly suggestive, but additional data at the level of specific type of breach will be needed before this model can be explored further. The occupancy model was unfortunately less successful, though it had the more ambitious objective of providing predictions across all the listed firms. The mean time to failure of the occupancy model was consistent with those from the hazard model, and we feel that with more extensive data collection, a more robust estimate could be obtained. We consider our estimates to be realistic but poorly supported by our current dataset for the occupancy structural model. We believe that with more information about breaches, that these model holds promise for successful future structural modeling.

Finally, our hypothesis $H_5$ tests of causality (counterfactuals) yielded equivocal results and we were unable determine the direction of causality. The only effect we found was with respect to audit fees which rose sharply from 2003 to about 2007 because of SOX compliance work, but dropped back down and dipped after the 2008 recession; these effects were driven by factors external to our study. This created a situation where the cumulative average returns pre- and post-breach were different for the audit fees, but did not specifically indicate a causal relationship.

This research provided insights into the value of SOX 404 and 302 attestations in managing the increasingly complex internal control environment that firms face. SOX 404 procedures are costly and controversial, but we have provided evidence to show that even in the limited realm of security breaches (only a small part of SOX attestation work) external and internal SOX audits provide important information that is not otherwise available to the firms and regulatory organizations. As with all empirical security research, ours is hampered by the fact that cybercrime is a low-risk high return vocation. Industry estimates suggest that only 3% to 30% of breaches are publicly reported in our research dataset (Menn 2012) and often breaches are not reported by firms because they may signal to customers and investors a lack of internal control. the best hackers are likely to be state-sponsored security professionals, using methods that are undetectable by current technologies and which can subvert internal controls and compromise assets without ever being detected. This is especially true with financial institutions such as banks and insurance companies, who can avoid some of the public scrutiny of customer facing retail and Internet firms. With clever use of existing breach data for construction of structural models, and more powerful tests, we are confident that these limitations in empirical research on control breaches can be overcome.